\documentclass[12pt]{article} %[reprint,aps]{revtex4-1}
\usepackage{epsfig, amsmath, amssymb}
\hypersetup{bookmarksopen=true, bookmarksnumbered=true,
 pdfstartview={FitH}, pdfborder={0 0 0}, colorlinks=true,
 linkcolor=red, linktocpage=true, citecolor=blue, urlcolor=blue,
 unicode=true, pdftitle={Entangled Spins and Ghost Spins},
 pdfauthor={DJ,KN}, pdfsubject={Entanglement of positive and
   indefinite norm sectors}, pdfkeywords={dS, EE, b-c ghost system,
   toy model}}
\usepackage{cite}
\newcommand\arXivid[1] {\href{http://arxiv.org/abs/#1}{\tt arXiv:#1}}
\newcommand\jhep[3]{{\it JHEP\ }\href{http://inspirehep.net/search?ln=en&ln=en&p=find+j+''JHEP,#1,#3''&of=hb&action_search=Search&sf=&so=d&rm=&rg=25&sc=0}
{{\bf #1} (#2) #3}}

\newcommand\prl[3] {{\it Phys.\ Rev.\ Lett.\
  }\href{http://dx.doi.org/10.1103/PhysRevLett.#1.#3}{{\bf #1} (#2) #3}}
\newcommand{\be}{\begin{equation}}
\newcommand{\ee}{\end{equation}}
\newcommand{\ben}{\begin{equation}}
\newcommand{\een}{\end{equation}}
\newcommand{\bea}{\begin{eqnarray}}
\newcommand{\eea}{\end{eqnarray}}
\newcommand{\bA}{\begin{array}}
\newcommand{\eA}{\end{array}}
\newcommand{\bc}{\begin{center}}
\newcommand{\ec}{\end{center}}
\newcommand{\al}{\alpha}

\newcommand{\ra}{\rightarrow}
\newcommand{\del}{\partial}

\newcommand{\ie}{{\it i.e.}}
\newcommand{\eg}{{\it e.g.}}

\newcommand{\ua}{\uparrow}
\newcommand{\da}{\downarrow}
\newcommand{\lan}{\langle}
\newcommand{\ran}{\rangle}
\newcommand{\ura}{\underrightarrow}
\newcommand{\ula}{\underleftarrow}

\numberwithin{equation}{section}

\begin{document}

%\ifeprint
%\fi

\begin{titlepage}
%\vspace{25mm}

\bc

%%\hfill  {TIFR/TH/09-12} \\
\hfill % {\tt arXiv:0909.4731 [hep-th]} 
\\         [20mm]
%%X\vfill

%% 
{\Huge Ghost-spin chains, entanglement \\ [2mm] 
and $bc$-ghost CFTs}
\vspace{10mm}

{\large Dileep P.~Jatkar$^{1,2}$ and K.~Narayan$^3$} \\
\vspace{3mm}
{\it 1. Harish-Chandra Research Institute \\
Chhatnag Road, Jhusi, Allahabad 211019, India\\[2mm]
2. Homi Bhabha National Institute\\
Training School Complex, Anushakti Nagar, Mumbai 400085, India\\[2mm]
3. Chennai Mathematical Institute, \\
SIPCOT IT Park, Siruseri 603103, India.\\[4mm]}
%% {\small Email: \ dileep@hri.res.in, narayan@cmi.ac.in}\\
%% 
\ec
%% %\medskip
\vspace{20mm}

\begin{abstract}
We study 1-dimensional chains of ghost-spins with nearest neighbour
interactions amongst them, developing further the study of ghost-spins 
in previous work, defined as 2-state spin variables with indefinite
norm. First we study finite ghost-spin chains with Ising-like nearest
neighbour interactions: this helps organize and clarify the study of
entanglement earlier and we develop this further. Then we study a
family of infinite ghost-spin chains with a different Hamiltonian
containing nearest neighbour hopping-type interactions. By defining
fermionic ghost-spin variables through a Jordan-Wigner transformation,
we argue that these ghost-spin chains lead in the continuum limit to
the $bc$-ghost CFTs.
\end{abstract}

\end{titlepage}

%\begin{flushright} {\tiny ( \end{flushright}

%\newpage %{\tiny 
{\footnotesize
\begin{tableofcontents}
\end{tableofcontents}}
%%}

%\vspace{5mm}

\section{Introduction} \label{sec:introduction}

Theories with gauge symmetry described in a covariant formulation are
known to contain sectors with negative norm states in part described
by ghost field excitations, although the physical content is often
captured by a physical positive norm subspace alone. In 2-dim
conformal field theories, ghost sectors have negative central charge,
a reflection of the negative norm states.
In \cite{Narayan:2016xwq}, the entanglement entropy properties of
certain 2-dim ghost conformal field theories were studied, with the
finding that entanglement entropy was non-positive under certain 
conditions: we will discuss later the motivations that led to those
investigations. Ghost-spins were constructed as a simple quantum
mechanical toy model for theories with negative norm states and a
study of their entanglement properties was also carried out.  This was
developed further in \cite{Jatkar:2016lzq} where ensembles of
ghost-spins entangled with ordinary spins was studied in more detail.
While a single spin is defined as a 2-state spin variable with a
positive definite inner product
$\langle \uparrow|\uparrow\rangle = 1 = \langle \downarrow|\downarrow\rangle$ 
and 
$\langle \uparrow|\downarrow\rangle = 0 = \langle \downarrow|\uparrow\rangle$,
a single ghost-spin is defined as a 2-state spin variable with the 
indefinite inner product
$\langle \uparrow|\uparrow\rangle = 0 = \langle \downarrow|\downarrow\rangle$ 
and 
$\langle \uparrow|\downarrow\rangle = 1 = \langle \downarrow|\uparrow\rangle$,\
akin to the inner products in the $bc$-ghost system as is well-known
(see \eg\ \cite{polchinskiTextBk}). The indefinite norm leads to
negative norm states: tracing over a subset of these leads to a
reduced density matrix for the remaining variables that is not
positive definite, and thereby to non-positive entanglement entropy
(EE) (reviewed in sec.~\ref{sec:revsGs}). This study involves only
information about the state of the system, with no recourse to
dynamics.

In this work, we will study dynamical models of ensembles of
ghost-spins.  In sec.~3, we study 1-dimensional ghost-spin chains with
a finite number of ghost-spins and discuss certain Ising-like nearest
neighbour interactions. This helps organize the study of ghost-spin
entanglement in \cite{Narayan:2016xwq,Jatkar:2016lzq}. In particular
we describe the properties of the reduced density matrix and
entanglement in such systems.

In sec.~4, we study a concrete example of a family of infinite
ghost-spin chains motivated by the well-known family of $bc$-ghost
conformal field theories. The $bc$-ghost system has been discussed
extensively in \eg\ \cite{polchinskiTextBk,Blumenhagen:2013fgp}, as
well as \cite{Friedan:1985ge}, and more recently \cite{Saleur:1991hk,
  Kausch:1995py,Kausch:2000fu,Flohr:2001zs,Krohn:2002gh}: they arise
as Fadeev-Popov ghosts under gauge fixing, as is well-known from the
$c=-26$ $bc$-CFTs in worldsheet string theory. The $bc$-ghost CFT with
$c=-2$ can also be thought of as the nonlogarithmic subsector of
$c=-2$ logarithmic CFTs consisting of 2-dim anticommuting (ghost)
scalars, \eg\ \cite{Gurarie:1993xq,Gurarie:1997dw,Gurarie:2013tma,
  Kausch:1995py,Kausch:2000fu,Flohr:2001zs,Krohn:2002gh}. By
constructing fermionic ghost-spin variables through a version of the
Jordan-Wigner transformation, we show that the infinite ghost-spin
chains here in fact lead to the $bc$-ghost CFTs in the continuum
limit. Our investigation here is motivated by the well-known that an
Ising spin chain in the continuum limit maps to a conformal field
theory of free massless fermions (see \eg\
\cite{Kogut:1979wt},\cite{SachdevQPT}).  Sec.~5 contains a Discussion.

\section{Reviewing ghost-spin ensembles and entanglement}
\label{sec:revsGs}

Here we review ``ghost-spins'' constructed in \cite{Narayan:2016xwq}, 
ensembles of ghost-spins and spins studied in \cite{Jatkar:2016lzq}, 
and their entanglement structures.

Firstly, for ordinary spin variables with a 2-state Hilbert space 
consisting of\ $\{ \uparrow,\ \downarrow \}$, we take the usual positive 
definite norms in the Hilbert space\ 
\be
{\rm spins}:\qquad\qquad\qquad \langle \uparrow | \uparrow\rangle = 
\langle \downarrow | \downarrow\rangle = 1\ ,\qquad 
\langle \uparrow | \downarrow\rangle = 
\langle \downarrow | \uparrow\rangle = 0\ .
\ee
Then a generic state 
$|\psi\rangle = c_1 | \uparrow\rangle + c_2 | \downarrow\rangle$ has 
adjoint $\langle\psi| = c_1^* \langle\uparrow| + c_2^* \langle\downarrow|$ 
and a positive definite norm $\langle \psi| \psi\rangle = |c_1|^2 + |c_2|^2$.
Thus states can be normalized as $\langle \psi| \psi\rangle = 1$.\ 
For 2-spin systems, entangled states $|\psi\ran = \psi^{ij}|ij\ran$ lead 
after tracing over say the second spin to a reduced density matrix with 
components\ $\rho_A^{ij} = \delta_{kl}\psi^{ik}(\psi^*)^{jl}$ which is 
automatically normalized as $tr\rho_A=\lan\psi|\psi\ran=1$. The positive 
definite norm structure here ensures that $\rho_A$ has von Neumann entropy 
$S_A = -tr \rho_A \log \rho_A = -\sum_i \rho_A(i) \log \rho_A(i)$ 
which is positive definite since each eigenvalue $0<\rho_A(i)<1$ makes 
the $-\log\rho_A(i) > 0$.

We define a single ``ghost-spin'' by a similar 2-state Hilbert space\ 
$\{ \uparrow,\ \downarrow \}$, but with norms 
\be\label{ghost-norms}
{\rm ghost\ spins}:\qquad\qquad\qquad \langle \uparrow |
\uparrow\rangle = 
\langle \downarrow | \downarrow\rangle = 0\ ,\qquad\quad
\langle \uparrow | \downarrow\rangle = 
\langle \downarrow | \uparrow\rangle = 1\ .
\ee
This is akin to the normalizations in the $bc$-ghost system in
\cite{Narayan:2016xwq} (see \eg\ \cite{polchinskiTextBk}, Appendix, 
vol.~1 where this inner product appears). Now a generic state and 
its non-positive norm are
\be\label{<psi|psi>}
|\psi\rangle = \psi^{\ua} |\uparrow\rangle + \psi^{\da} |\downarrow\rangle 
\qquad \Rightarrow\qquad \langle \psi| \psi \rangle = \gamma_{\al\beta}
\psi^\al (\psi^\beta)^* = \psi^{\ua} (\psi^{\da})^* + 
(\psi^{\ua})^* \psi^{\da}\ ,
\ee
where the adjoint is 
$\langle\psi| = (\psi^{\ua})^* \langle\ua| + (\psi^{\da})^* \langle\da|$, 
and the ghost-spin inner product is given by the off-diagonal metric 
$\gamma_{\ua\da}=\gamma_{\da\ua}=1$.\ An alternative convenient basis is
\be\label{ghostbasis+-}
\begin{split}
&|\pm\rangle \equiv {1\over \sqrt{2}} \big(|\uparrow\rangle\ \pm\
|\downarrow\rangle \big)\ ,
\quad 
\langle +| + \rangle = \gamma_{++} = 1\ ,\\ 
&\langle -| - \rangle  = \gamma_{--} = -1\ ,
\quad\quad \langle +| - \rangle = \langle -| + \rangle = 0\ .
\end{split}
\ee
A generic state with nonzero norm can be normalized to norm $+1$ or $-1$. 
Then a negative norm state can be written as\ 
$|\psi\rangle = \psi^+ |+\rangle + \psi^- |-\rangle$ with 
$\langle \psi| \psi\rangle = |\psi^+|^2 - |\psi^-|^2 = -1$.
For a system of two ghost-spins,\
$|s_A s_B\rangle \equiv
\{|\uparrow \uparrow\rangle ,  |\uparrow \downarrow\rangle , 
|\downarrow \uparrow\rangle ,  |\downarrow \downarrow\rangle \} 
\equiv \{ | + + \rangle , | + - \rangle , | - + \rangle , 
| - - \rangle\} $   %\ee
are basis states.
We define the states, adjoints and (indefinite) norms as
\be\label{2gsnorms}
|\psi\rangle = \sum \psi^{\alpha\beta} |\alpha\beta\rangle ,\ \ \
\langle\psi| = \sum \langle \alpha\beta| {\psi^{\alpha\beta}}^* ,
\qquad \langle\psi|\psi\rangle\ = \langle \kappa| \alpha\rangle \langle 
\lambda| \beta{}\rangle \psi^{\alpha\beta} {\psi^{\kappa\lambda}}^* \equiv 
\gamma_{\alpha\kappa} \gamma_{\beta\lambda} \psi^{\alpha\beta} 
{\psi^{\kappa\lambda}}^*\ 
\ee
where repeated indices, as usual, are summed over.
A generic normalized positive/negative norm state 
$|\psi\rangle =  \psi^{++} | + + \rangle + \psi^{+-} | + - \rangle 
+ \psi^{-+} | - + \rangle + \psi^{--} | - - \rangle$ has norm
\be\label{psiijNorm1}
\langle\psi| \psi\rangle = |\psi^{++}|^2 + |\psi^{--}|^2 
- |\psi^{+-}|^2 - |\psi^{-+}|^2 = \pm 1\ ,
\ee
normalized with norm $\pm 1$, using the diagonal metric
(\ref{ghostbasis+-}) in the $|\pm\rangle$ basis.

The density matrix for the full system is
$\rho=|\psi\rangle\langle\psi|  
= \sum \psi^{\alpha\beta} {\psi^{\kappa\lambda}}^* 
|\alpha\beta \rangle\langle \kappa\lambda|$.  We split this sytem into
two subsystems $A$ and $B$.  The reduced density matrix for the 
subsystem $A$, which consists of one ghost-spin is obtained by carrying 
out the trace over the subsystem $B$ (environment) consisting of the 
other ghost-spin. This process can be defined on a multi-spin state 
using a partial contraction as 
\be\label{2gsrdm}
\rho_A = tr_B \rho \equiv (\rho_A)^{\alpha\kappa} |\alpha\rangle \langle \kappa|\ ,\qquad\ \  
(\rho_A)^{\alpha\kappa}\ =\ \gamma_{\beta\lambda} \psi^{\alpha\beta} {\psi^{\kappa\lambda}}^*\ =\ 
\gamma_{\beta\beta}  \psi^{\alpha\beta} {\psi^{\kappa\beta}}^*\ ,
\ee
\bea\label{rhoAc1234}
\Rightarrow\qquad\qquad 
(\rho_A)^{++} =\ |\psi^{++}|^2 - |\psi^{+-}|^2\ , &\quad&
(\rho_A)^{+-} =\ \psi^{++} {\psi^{-+}}^* - \psi^{+-} {\psi^{--}}^*\ , 
\nonumber\\
(\rho_A)^{-+} =\ \psi^{-+} {\psi^{++}}^* - \psi^{--} {\psi^{+-}}^*\ , &\quad& 
(\rho_A)^{--} =\ |\psi^{-+}|^2 - |\psi^{--}|^2\ , 
\eea
Then $tr \rho_A = \gamma_{\alpha\kappa} (\rho_A)^{\alpha\kappa} =
(\rho_A)^{++} - (\rho_A)^{--}$. 
Thus the reduced density matrix is normalized to have\ 
$tr \rho_A = tr \rho = \pm 1$ depending on whether the state 
(\ref{psiijNorm1}) is positive or negative norm.
The entanglement entropy calculated as the von Neumann entropy of 
$\rho_A$ is
\be\label{EErhoA}
S_A = -\gamma_{\alpha\beta} (\rho_A \log \rho_A)^{\alpha\beta}\ =\ 
- \gamma_{++} (\rho_A \log \rho_A)^{++}
- \gamma_{--} (\rho_A \log \rho_A)^{--}
\ee
where the last expression pertains to the $|\pm\rangle$ basis with
$\gamma_{\pm\pm}=\pm 1$.  This requires defining $\log\rho_A$ as usual
as an operator expansion: here the contractions use the indefinite
norm and are perhaps more transparent in terms of the mixed-index
reduced density matrix $(\rho_A)^\alpha{_\kappa}$.

As an illustration, consider a simple family of states studied in
\cite{Narayan:2016xwq} with a diagonal reduced density matrix, so
$\log\rho_A$ is also diagonal and easily calculated. For the states
(\ref{psiijNorm1}), this gives
${\psi^{-+}}^* = {\psi^{+-} {\psi^{--}}^*\over \psi^{++}}$ and
$\langle\psi| \psi\rangle = (|\psi^{++}|^2 - |\psi^{+-}|^2) 
(1 + {|\psi^{--}|^2\over |\psi^{++}|^2} ) = \pm 1$: so
(\ref{rhoAc1234}) gives
\bea\label{Ex:2gsEE}
&& (\rho_A)^{\al\beta}|\al\ran\lan\beta| = \pm x |+ \rangle\langle +|\
\mp\ (1-x) |- \rangle\langle -|\ , \qquad
x = {|\psi^{++}|^2\over |\psi^{++}|^2 + |\psi^{--}|^2} \qquad [0 < x < 1] ,
\nonumber\\
&& (\rho_A)_\alpha^\kappa = \gamma_{\alpha\beta} (\rho_A)^{\beta\kappa} :
\qquad (\rho_A)^+_+ = \pm x\ ,\qquad (\rho_A)^-_- = \pm (1-x)\ ,\quad\
\eea
where the $\pm$ pertain to positive and negative norm states respectively.
The location of the negative eigenvalue is different for positive and 
negative norm states, leading to different results for the von Neumann 
entropy. For negative norm states, $(\rho_A)^{++} < 0,\ (\rho_A)^{--} > 0$.\
From the mixed-index RDM in the second line above, we see that\ 
$tr\rho_A = (\rho_A)^+_+ + (\rho_A)^-_- = \pm 1$ manifestly. Now we
obtain\  
$(\log\rho_A)^+_+ = \log (\pm x)$ and $(\log\rho_A)^-_- = \log (\pm (1-x))$,
the $\pm$ referring again to positive/negative norm states respectively. 
The entanglement entropy (\ref{EErhoA}) becomes\
$S_A = - (\rho_A)^+_+ (\log\rho_A)^+_+ - (\rho_A)^-_- (\log\rho_A)^-_-$\ 
and so
\bea\label{Ex:2gsEE2}
\langle\psi| \psi\rangle > 0: &&  S_A = - x\log x - (1-x) \log (1-x) > 0\ ,\\
\langle\psi| \psi\rangle < 0: && S_A =  x\log (-x) + (1-x) \log
(-(1-x)) \nonumber \\
&&\phantom{S_A}= x\log x + (1-x) \log (1-x) + i\pi(2n+1)x + i\pi(2m+1)(1-x)\ .
\nonumber
\eea
For positive norm states, $S_A$ is manifestly positive since $x<1$,
just as in an ordinary 2-spin system.\ For negative norm states, we
note that for the principal branch, \ie\ $n=m$, the imaginary part is
independent of $x$: in other words the imaginary part is the same for
all such negative norm states provided we choose the same branch of
the logarithm.  In what follows whenever we get a logarithm with
negative argument we will list all branches but in our analysis we
will consider the principal branch only (with $n,m=0$), \ie\ we will 
effectively set $\log (-1)=i\pi$. The real part of entanglement
entropy is negative since $x<1$ and the logarithms are negative.

We now review ensembles of ghost-spins and spins, possibly entangled,
regarding them in general as toy models for quantum systems containing
negative norm states. For multiple variables, the spin Hilbert space has 
a positive definite metric $g_{ij}=\delta_{ij}$, while the ghost-spin 
states have a non-positive metric $\gamma_{ij}$, with components\ 
$\gamma_{++}=1,\ \gamma_{--}=-1$, as in (\ref{ghostbasis+-}) by a basis change\ 
$\{|\uparrow\rangle,|\downarrow\rangle\}\ra \{|+\rangle,|-\rangle\}$\ 
which makes negative norm states manifest. The entanglement entropy 
properties of the reduced density matrix after tracing over
ghost-spins in such systems was studied in \cite{Jatkar:2016lzq}.

In general, the Hilbert space of spins and ghost-spins contains
positive as well as negative norm states. One might ask if the
entanglement entropy $S_A$ of $\rho^s_A$ is uniformly positive for all
positive norm states, and uniformly negative for all negative norm
states.  This can be shown to be identically true when the spin sector
is not entangled with the ghost-spin sector (both of which could be
entangled within themselves).
Firstly, considering observables of the spin variables alone, we expect 
that the correlation function satisfies 
$\langle \psi| O_s|\psi\rangle = tr_s (O_s\rho^s)$. Performing the trivial 
trace over all the ghost-spins shows that the reduced density matrix 
for the remaining spin sector alone is $\rho^s = tr_{gs} \rho$. 
Now disentangled ghost-spins and spins can be represented as product 
states with
\bea
|\psi\rangle = |\psi_s\rangle\ |\psi_{gs}\rangle , && \!\!
\langle\psi|\psi\rangle = \langle\psi_s|\psi_s\rangle\
\langle\psi_{gs}|\psi_{gs}\rangle\ ,\\
\langle\psi_s|\psi_s\rangle = g_{i_1j_1} \ldots g_{i_nj_n} 
(\psi_s)^{i_1i_2\ldots} (\psi_s)^{j_1j_2\ldots *} > 0 , && \!\!
\langle\psi_{gs}|\psi_{gs}\rangle = \gamma_{i_1j_1} \ldots \gamma_{i_nj_n} 
(\psi_{gs})^{i_1i_2\ldots} (\psi_{gs})^{j_1j_2\ldots *},\nonumber
\eea
where the spin inner product is positive definite while 
$\langle\psi_{gs}|\psi_{gs}\rangle$ can be positive or negative.
Normalizing positive/negative norm states to have norm $\pm 1$ 
respectively gives
\be
\langle\psi_{gs}|\psi_{gs}\rangle \gtrless 0\quad \Rightarrow\quad
\langle\psi|\psi\rangle = \langle\psi_s|\psi_s\rangle\ 
\langle\psi_{gs}|\psi_{gs}\rangle = \pm 1 \qquad\quad
[\langle\psi_s|\psi_s\rangle > 0]\ .
\ee
The reduced density matrix after tracing over all ghost-spins is
$\rho_A^s = tr_{gs} \big( |\psi_s\rangle\ |\psi_{gs}\rangle \langle\psi_s|\ 
\langle\psi_{gs}| \big)$\ giving
\be
(\rho_A^s)^{i_1\ldots , k_1\ldots} = \langle\psi_{gs}|\psi_{gs}\rangle\ 
(\psi_s)^{i_1\ldots} (\psi_s)^{k_1\ldots *}
= \pm {1\over \langle\psi_s|\psi_s\rangle} 
(\psi_s)^{i_1\ldots} (\psi_s)^{k_1\ldots *}\ .
\ee
This implies the normalization\ $tr \rho_A^s = \pm 1$ for 
positive/negative norm states\ 
$\big(\langle\psi|\psi\rangle \gtrless 0\big)$.

We see that the sign of the norm of the state enters as an overall
sign in $\rho^s_A$. Thus for positive norm states, $\rho_A^s$ is
positive definite with eigenvalues\ $0<\lambda_i< 1$\ satisfying\
$\sum_i\lambda_i=1$ giving positive definite entanglement entropy\
$S_A = - {\rm tr}_s\ \rho_A^s \log\rho_A^s = -\sum_i\lambda_i\log\lambda_i > 0$.
For negative norm states however, we see that $\rho_A^s$ is negative 
definite with eigenvalues\ $-\lambda_i$. Thus the von Neumann entropy is\ 
$S_A = - {\rm tr}_s\ \rho_A^s \log\rho_A^s 
= -\sum_i (-\lambda_i)\log(-\lambda_i) = \sum_i \lambda_i\log\lambda_i 
+ i\pi$\ (taking $\log(-1)=i\pi$ as stated earlier). The entanglement 
entropy thus has a negative definite real part and a constant imaginary 
part, similar to the subfamily (\ref{Ex:2gsEE}), (\ref{Ex:2gsEE2}), 
of two ghost-spin states.

When the spins are entangled with the ghost-spins, then this
straightforward correlation between positive norm states and
positivity of the entanglement entropy appears to not be true. With
$\rho_A$ the reduced density matrix for the remaining spin variables
after tracing over all the ghost-spins, the von Neumann entropy
contains components of $\rho_A$ which in turn contains linear
sub-combinations of the norm of the state. Thus even for positive norm
states, some components of $\rho_A$ can be negative in general (while
keeping positive the trace of $\rho_A$, which is the norm of the
state): this leads to new entanglement patterns in general. Requiring
that positive norm states give positive entanglement $S_A$ amounts to
requiring that the components $(\rho_A)^{IJ}$ are positive ($I,J$
being labels for the remaining spin variables): this is only true for
specific subregions of the Hilbert space, \ie\ only certain families
of states.
When the number of ghost-spins is even, we can restrict to subfamilies
of states which have correlated ghost-spins, \ie\ the ghost-spin
values are the same in each basis state. This implies that all allowed
states are positive norm, \ie\ negative norm states are excluded. This
restricts to half the space of states which are now all positive norm,
and the entanglement entropy is manifestly positive. The intuition
here is in a sense akin to simulating \eg\ the $X^{\pm} + bc$
subsector of the 2-dim sigma model representing the string worldsheet
theory: in general negative norm states are cancelled between
$X^{\pm}$ and the $bc$-ghost subsectors in the eventual physical
theory. The more general subsectors in the Hilbert space where
$\rho_A$ gives positive entanglement entropy for positive norm states
can then be interpreted as the component of the state space that is
connected to this correlated ghost-spin sector. As an example, 
consider a system of one spin entangled with two ghost-spins: the 
general state is\ $\psi^{i,\al\beta} |i\rangle|\al\beta\rangle$ and 
tracing over both ghost-spins leads to the reduced density matrix
$(\rho_A)^{ik} = \gamma_{\al\sigma} \gamma_{\beta\rho} 
\psi^{i,\al\beta} (\psi^*)^{k,\sigma\rho}$. The subfamily of states 
represented by\
$|\psi\rangle = \psi^{+,++} |+\rangle |++\rangle 
+ \psi^{+,--} |+\rangle |--\rangle 
+\ \psi^{-,++} |-\rangle |++\rangle + \psi^{-,--} |-\rangle |--\rangle$
characterizes here the subspace of correlated ghost-spins:
this is manifestly positive norm so that $\rho_A$ is uniformly 
positive definite as is the entanglement entropy. This is also true 
for part of the component of the Hilbert space continuously connected 
to this subspace. For instance, the family of states
$|\psi\rangle = \psi^{+,++} |+\rangle |++\rangle 
+ \psi^{+,+-} |+\rangle |+-\rangle 
+\ \psi^{-,-+} |-\rangle |-+\rangle + \psi^{-,--} |-\rangle |--\rangle$
have norm $\langle\psi|\psi\rangle = |\psi^{+,++}|^2 - |\psi^{+,+-}|^2
- |\psi^{-,-+}|^2 + |\psi^{-,--}|^2$ and lead to a diagonal reduced 
density matrix\ $(\rho_A)^{++} = |\psi^{+,++}|^2 - |\psi^{+,+-}|^2,\
(\rho_A)^{--} = -|\psi^{-,-+}|^2 + |\psi^{-,--}|^2$. This is positive 
definite as long as the states are ``mostly'' correlated ghost-spins, 
\ie\ the components $\psi^{+,+-}, \psi^{-,-+}$, are appropriately small.
More generally, even ghost-spins allow sensible interpretations.

For systems with odd number of ghost-spins however, such a consistent
subfamily of correlated ghost-spin states does not exist so it is not
possible to uniformly pick a family of entangled states mentioned
above such that positive norm states give positive entanglement
entropy.
For example, with one spin entangled with one ghost-spin, the general 
state is $\psi^{i,\al} |i\rangle |\al\rangle$ giving 
$(\rho_A)^{ik} = \gamma_{\al\beta} \psi^{i,\al} (\psi^*)^{k,\beta}$ as 
the reduced density matrix. A simple entangled state is
$|\psi\rangle = \psi^{+,+} |+\rangle|+\rangle + 
\psi^{-,-} |-\rangle|-\rangle$ with 
$(\rho_A)^{++} = |\psi^{+,+}|^2,\ (\rho_A)^{--} = - |\psi^{-,-}|^2$ and
$|\psi^{+,+}|^2 - |\psi^{-,-}|^2 = \pm 1$. Thus 
$(\log\rho_A)^+_+ = \log(|\psi^{+,+}|^2) ,\ 
(\log\rho_A)^-_- = \log(-|\psi^{-,-}|^2)$, so the entanglement entropy 
is\ $S_A = -|\psi^{+,+}|^2 \log \big(|\psi^{+,+}|^2\big) 
+ |\psi^{--}|^2 \log \big(|\psi^{-,-}|^2\big) + |\psi^{-,-}|^2 (i\pi)$. 
Thus a positive norm state does not give positive EE. Likewise, for 
a system of $n$ ghost-spins with $n$ odd (and no spins), the family of 
states $|\psi\rangle = \psi^{++\ldots} |++\ldots\rangle 
+ \psi^{--\ldots} |--\ldots\rangle$ with norm\ 
$\langle\psi|\psi\rangle = |\psi^{++\ldots}|^2 + (-1)^n |\psi^{--\ldots}|^2$ 
leads to a reduced density matrix\
$(\rho_A)^+_+ = (\rho_A)^{++} = |\psi^{++\ldots}|^2 ,\ \ 
(\rho_A)^-_- = -(\rho_A)^{--} = (-1)^n |\psi^{--\ldots}|^2$, structurally 
similar to the one spin and one ghost-spin case above. That is, there 
always exist positive norm states that lead to entanglement entropy with 
negative real part (and nonzero imaginary part).

In the next section, we will discuss how introducing nearest neighbour
interactions in the context of a finite ghost-spin chain organizes our
understanding of this subspace of correlated ghost-spins and small 
deformations thereof.

\section{Interactions and finite ghost-spin chains}

We study 1-dimensional chains with a finite number of ghost-spins in
this section: we imagine this to be a generalization to ghost-spins of
ordinary 1-dimensional spin chains that are familiar in statistical
physics and condensed matter systems (see \eg\ \cite{Kogut:1979wt},
\cite{SachdevQPT}).
The simplest such configuration here consists of two ghost-spins: 
consider an Ising-like nearest neighbour interaction
\be\label{2gsIsingHint}
H = -J s s'\ ,\qquad\qquad s|\pm\rangle=\pm|\pm\rangle\ ,
\ee
where $s, s'$ are ghost-spin variables and we have written their action 
in the $\{+-\}$ basis. If $J>0$, this has the same structure as for 
ordinary spin ferromagnetic interactions.
For instance, 
\be
H|\pm \pm \rangle = -J|\pm \pm \rangle ,\qquad 
H|\pm \mp \rangle = +J|\pm \mp \rangle .
\ee
The expectation values are the same as for spins,
\be
\langle H\rangle_{\pm\pm} = {\langle \pm\pm| H |\pm\pm\rangle\over 
\langle \pm\pm| \pm\pm\rangle} = -J\ ,\qquad
\langle H\rangle_{\pm\mp} = {\langle \pm\mp| H |\pm\mp\rangle\over 
\langle \pm\mp| \pm\mp\rangle} = +J\ ,
\ee
where \eg\ $\langle \pm\mp| \pm\mp\rangle = -1$ using the norms in 
(\ref{2gsnorms}), and the minus sign cancels in the numerator and 
denominator in the expectation value\ (note that 
$\langle \pm\mp| H |\pm\mp\rangle = +J (-1)$,\ \ie\ these 
correlation functions acquire an additional minus sign).\ 

The above nearest neighbour interaction implies that two
positive/negative norm configurations attract while one positive and
one negative norm repel.
This suggests the mapping
\be
[{\rm ghost\ spin}]\ \ \ \{ +, -\}\quad\ \equiv\quad \{ \uparrow, \downarrow\}
\ \ \ [{\rm spin}]
\ee
so that the ghost-spin ensemble with the interactions as defined 
here in the $\{ +, - \}$-basis maps identically to an ordinary spin 
ensemble in the $\{ \uparrow, \downarrow \}$-basis.
Thus we have
\bea
{\rm ground\ states}: && |++\rangle,\ \ |--\rangle\ ,\qquad\qquad 
\langle\pm\pm| \pm\pm\rangle = +1\ ,\nonumber\\
{\rm excited\ states}: && |+-\rangle,\ \ |-+\rangle\ ,\qquad\qquad
\langle\pm\mp| \pm\mp\rangle = -1\ ,
\eea
\ie\ the ground states are positive norm while the excited states 
are negative norm.
The partition function is
\be\label{2gsZ2}
Z = \sum_n e^{-\beta E_n} = 2\big( e^{-\beta J} + e^{\beta J} \big) 
\equiv 2 Z_{(2)}\ ,
\ee
identical to that for two Ising-like spins, as expected from the 
mapping $\{ +, -\} \equiv \{ \uparrow, \downarrow\}$. This is despite 
the fact that the ghost-spin system has negative norm states. In 
this regard, we should note that this is akin to the partition 
function for the $bc$-ghost CFT with $c=-2$ which is positive definite 
although there is a plethora of negative norm states.

\vspace{2mm}

\noindent {\bf Time evolution}:\ \ 
Let us now study the time evolution generated by this Hamiltonian 
(\ref{2gsIsingHint}) using the usual rules of quantum mechanics. 
It is clear that all eigenstates 
$H|\Psi_E\rangle = E|\Psi_E\rangle$ evolve simply through phases so that 
\ $|\Psi(t)_E\rangle = e^{-iHt}|\Psi(0)_E\rangle = e^{-iEt}|\Psi(0)_E\rangle$ 
in the Schrodinger picture. This implies that
\be
\begin{split}
  \left( |++\rangle (t),\ |--\rangle (t) \right)&= e^{-i(-J)t}\left( |++\rangle (0), |--\rangle \right)
(0)\ ,\\
\left( |+-\rangle (t),\ |-+\rangle (t)  \right) &= e^{-i(+J)t}\left( |+-\rangle (0), |-+\rangle (0)  \right)\ .
\end{split}
\ee
Then a generic state evolves as\
$|\psi(t)\rangle = c_1 e^{+iJt} |++\rangle + c_2 e^{-iJt} |+-\rangle 
+ c_3 e^{-iJt} |-+\rangle + c_4 e^{+iJt} |--\rangle$,\
and the norm\ 
$\langle\psi(t)|\psi(t)\rangle = |c_1|^2 + |c_4|^2 - |c_2|^2 - |c_3|^2 
= \langle\psi(0)|\psi(0)\rangle = \pm 1$\
remains invariant under time evolution. The phases cancel out since 
the basis states are $H$-eigenstates and orthogonal to each other.
This means that $\pm$ve norm states evolve to $\pm$ve norm states and 
do not mix.

To contrast the present case of negative norm states with ordinary
spins, it is interesting to ask if a probabilistic interpretation exists. 
The amplitude for the state $|\Psi(0)\rangle$ 
to evolve to itself is given by $\langle\Psi(0)|e^{-iHt}|\Psi(0)\rangle$ 
which is
\be
\langle \Psi(0)|\Psi(t)\rangle = \left( |c_1|^2 + |c_4|^2 \right) e^{iJt} 
- \left( |c_2|^2 + |c_3|^2 \right) e^{-iJt} =
\left( |c_1|^2 + |c_4|^2 \right) \left( e^{iJt} - e^{-iJt} \right) 
\pm e^{-iJt}
\ee
using the norm condition above. By comparison, for ordinary spins, we
have\ $\sum_i|c_i|^2=1$ and\ $\langle \Psi(0)|\Psi(t)\rangle = 
( |c_1|^2 + |c_4|^2 ) ( e^{iJt} - e^{-iJt} ) +  e^{-iJt}$. So for positive 
norm states, the overlap amplitude for ghost-spins is of the same form 
as for ordinary spins: for negative norm states, the sign is different.

Consider now a state $|\Psi(0)\rangle=c_1|++\rangle+c_2|+-\rangle$ 
normalized as $|c_1|^2-|c_2|^2=\pm 1$:\ then the probabilities to be 
measured in $|++\rangle$ or $|+-\rangle$ are
\be
P_\Psi(++) = |\langle ++| \Psi(t)\rangle|^2 = |c_1|^2 > 0\ ,\qquad
P_\Psi(+-) = |\langle +-| \Psi(t)\rangle|^2 = |c_2|^2 = |c_1|^2 \mp 1\ .
\ee
Thus the total probability which is the sum of component probabilities
$P(++)+P(+-)$ is not unity, even when $|\Psi\ran$ is positive norm: 
probability conservation does not hold, since the negative norm 
components give a minus sign as expected (by comparison, for ordinary 
spins, we have 
$P(\uparrow\uparrow)+P(\uparrow\downarrow) = |c_1|^2+|c_2|^2=1$, with 
probability conserved as is familiar).

\vspace{2mm}

\noindent {\bf 3 ghost-spins:}\ \ The Hamiltonian for the ghost-spin chain is 
\be
H = -J\sum_{nn} ss' = -J s_1s_2 - J s_2s_3\ .
\ee
There are $2^3=8$ states $| \pm \pm \pm \rangle$ in all and their energies 
$E=\lan H\ran$ are
\bea
|+++\rangle,\ |---\rangle: &&\quad  E = -2J\ ,\nonumber\\
|+--\rangle,\ |++-\rangle,\ |--+\rangle,\ |-++\rangle: &&\quad 
E = +J-J = 0\ , \nonumber\\
|+-+\rangle,\ |-+-\rangle: &&\quad E = +2J\ .
\eea
It is clear that at each level, there are both positive and negative 
norm states: \eg\ at the ground state level, $|+++\rangle$ is positive 
norm while $|---\rangle$ is negative norm. This structure also holds 
for $N$ ghost-spins with $N$ odd, \ie\ the ground states contain 
$|--\ldots --\rangle$ which has negative norm.
The partition function is
\be
Z = 2 \big( e^{2\beta J} + 2 + e^{-2\beta J} \big) 
= 2 \big( e^{\beta J} + e^{-\beta J} \big)^2 = 2 Z_{(2)}^2\ ,
\ee
where $Z_{(2)}$ is the partition function (\ref{2gsZ2}) for 2 ghost-spins.
\vspace{2mm}

\noindent {\bf 4 ghost-spins:}\ \ The Hamiltonian for the ghost-spin chain is 
\be
H = -J\sum_{nn} ss' = -J s_1s_2 - J s_2s_3 - Js_3s_4\ .
\ee
There are $2^4=16$ states $| \pm \pm \pm \pm \rangle$ in all and the 
energies $E=\lan H\ran$ are
\bea\label{4gs<E>}
|++++\rangle,\ |----\rangle: &&\quad E = -3J\ ,\nonumber\\
|+++-\rangle,\ |++--\rangle,\ |+---\rangle,\qquad  && \nonumber\\
|-+++\rangle,\ |--++\rangle,\ |---+\rangle: &&\quad  E = -J\ ,
\nonumber\\
|+--+\rangle,\ |-++-\rangle,\ |--+-\rangle,\qquad && \nonumber\\
|++-+\rangle,\ |+-++\rangle,\ |+-++\rangle: &&\quad  E = +J\ ,
\nonumber\\
|+-+-\rangle, |-+-+\rangle: &&\quad E = +3J\ .
\eea
In this case, the ground states are uniformly {\bf positive norm}, as 
for two ghost-spins: these states fall in the category of ``correlated 
ghost-spins'' in \cite{Jatkar:2016lzq}. Some (but not all) of the 
excited states are negative norm. The partition function is
\be
Z = 2 \big( e^{3\beta J} + 3 e^{\beta J} + 3 e^{-\beta J} + e^{-3\beta J} \big)
= 2 \big( e^{\beta J} + 3 e^{-\beta J}\big)^3 = 2 Z_{(2)}^3\ ,
\ee
and is identical to the case of 4 Ising-like ordinary spins. This sort 
of structure persists for an even number of ghost-spins.

\vspace{2mm}

\noindent {\bf $N$ ghost-spins:}\ \ The Hamiltonian is 
\be\label{NgsIsingHint}
H = -J\sum_{nn} ss' = -J\sum_n s_ns_{n+1} 
= -Js_1s_2 - Js_2s_3 - \ldots - Js_{N-1}s_N\ .
\ee
There are $2^N$ states $|\pm^N\rangle$ in all.
If $N$ is even, the form of the ground states, and the corresponding 
energy, are
\be\label{NgsGndSt}
{\rm ground\ states}:\quad |+^N\rangle,\ |-^N\rangle, \qquad\quad 
E = -(N-1)J\ ,
\ee
and are both positive norm.\ 
(If $N$ is odd, then $|-^N\rangle$ has negative norm.)

Some of the excited states have negative norm, somewhat similar in
structure to the 4 ghost-spins case above. The highest energy states
(and corresponding energy) are of the form\ 
\be
|+-+-\ldots\rangle,\ |-+-+\ldots\rangle:\qquad E = (N-1)J\ ,
\ee
\ie\ maximally alternating $+, -$ ghost-spins (as in the case of 4
ghost-spins). These contain ${N\over 2}$ $\{-\}$ ghost-spins each (for 
$N$ even) and so are positive norm if ${N\over 2}$ is even.

The first excited level, with energy $\langle E\rangle = -(N-3)J$,
consists of states which have exactly one ``kink'' \ie\ one $\{+-\}$ 
(or one $\{-+\}$) interface, as illustrated above for 4 ghost-spins 
(\ref{4gs<E>})). In other words, (starting from the left) the first $-$ 
ghost-spin can be in one of $N-1$ locations out of $N$ (as in the 
second line in (\ref{4gs<E>})). Thus the first excited level comprises 
$2(N-1)$ states, of the form
\bea\label{Ngs1stExcSt}
&& |++\ldots +-\rangle ,\ \ |++\ldots +--\rangle ,\ \ \ldots\ \
|+--\ldots --\rangle, \nonumber\\
&& |-++\ldots ++\rangle ,\ \ |--++\ldots ++\rangle ,\ \ \ldots\ \
|---\ldots -+\rangle .
\eea
Higher excited states comprise multiple kinks and can be analysed similarly.
Note that a kink here has a single $+-$ or $-+$ interface and is in 
general distinct from a ``bulk'' flipped spin, since (in 1-dim) that 
would have two interfaces. We are considering
``open'' chains here: if we consider ``closed'' chains instead, then
the absence of endpoints means that each excitation of a flipped spin
comes with two kinks.

Thus the partition function has the form
\be\label{NgsZ}
Z = 2 \big( e^{(N-1)\beta J} + (N-1) e^{\beta (N-3) J} + \ldots + 
e^{-(N-1)\beta J} \big) 
= 2 \big( e^{\beta J} + e^{-\beta J} \big)^{N-1} = 2 Z_{(2)}^{N-1}\ ,
\ee
again a product over 2 ghost-spin partition functions (\ref{2gsZ2}).

Before discussing entanglement issues, we make a brief comment on the 
basis used here. Consider the action $s|\pm\rangle=\pm|\pm\rangle$ of the 
spin variable $s$: this means 
\be
s(|\uparrow\rangle \pm |\downarrow\rangle) = 
\pm (|\uparrow\rangle \pm |\downarrow\rangle) \quad\Rightarrow\qquad
s|\uparrow\rangle = |\downarrow\rangle\ , \quad 
s|\downarrow\rangle = |\uparrow\rangle\ .
\ee
Thus $s$ is like a spin-flip operator for these $\uparrow,\downarrow$ 
states, akin to the Pauli matrix $\sigma^x=(^0_1\ {}^1_0)$.
The Hamiltonian (\ref{2gsIsingHint}) itself, restricting to two 
ghost-spins for simplicity, can be written as
\be
H = -J s s' = -J \left(|++\rangle\langle ++| + |--\rangle\langle --| \right) 
+ J \left(|+-\rangle\langle +-| + |-+\rangle\langle -+| \right) \ .
\ee
Explicitly changing basis using
$|\pm\rangle = {1\over\sqrt{2}} 
\left(|\uparrow\rangle \pm |\downarrow\rangle\right)$ and expanding and 
simplifying $H$ gives
\be
H = -J \left(|\uparrow\uparrow\rangle\langle \downarrow\downarrow | 
+ |\downarrow\downarrow \rangle\langle \uparrow\uparrow|  
+  |\uparrow\downarrow\rangle\langle \downarrow\uparrow| 
+ |\downarrow\uparrow\rangle\langle \uparrow\downarrow| \right) \equiv 
-J \sigma_x \sigma_x'\ .
\ee
whereas $H \equiv -J \sigma_z \sigma_z'$ in the $\pm$-basis.

\subsection{Entanglement:\ $N$ ghost-spins}

For a chain of $N$ ghost-spins with Hamiltonian (\ref{NgsIsingHint}), the 
generic ground state and its norm are
\be\label{Ngndst}
|\psi_g\rangle = \psi_g^{+} |+^N\rangle + \psi_g^- |-^N\rangle\ ,
\qquad \langle \psi_g|\psi_g\rangle = |\psi_g^+|^2 + (-1)^N|\psi_g^-|^2\ ,
\ee
which is positive norm for $N$ even: the norm is\ 
$\langle \psi_g|\psi_g\rangle = |\psi_g^+|^2 + |\psi_g^-|^2 = 1$, after 
normalizing. This is an entangled state: tracing over $N-1$ ghost-spins, 
the reduced density matrix for the remaining single ghost-spin (using 
the notation in sec.~\ref{sec:revsGs}) has mixed index components\ 
$(\rho_A)^+_+ = |\psi_g^+|^2,\ (\rho_A)^-_- = |\psi_g^-|^2$, with 
von Neumann entropy\ $S_A = -(\rho_A)^+_+ \log (\rho_A)^+_+ - 
(\rho_A)^-_- \log (\rho_A)^-_- = -x\log x - (1-x)\log(1-x)>0$, where
$x\equiv |\psi_g^+|^2$ satisfies $0<x<1$.\ Thus the ground state 
entanglement entropy of the $N$ ghost-spin chain is manifestly 
positive definite, for $N$ even.

The excited states include negative norm states and these can 
exhibit new entanglement patterns. For instance for the case of 4 
ghost-spins above, consider from (\ref{4gs<E>}) the states
\bea\label{4gs+++-}
&& |\psi\rangle = \psi^{++++}|++++\rangle + \psi^{----}|----\rangle + 
\psi^{+++-}|+++-\rangle + \psi^{---+}|---+\rangle\ ,\ \nonumber\\
&& \langle\psi|\psi\rangle = |\psi^{++++}|^2 + |\psi^{----}|^2 
- |\psi^{+++-}|^2 - |\psi^{---+}|^2 = \pm 1\ ,
\eea
where we are normalizing positive/negative norm states to $\pm 1$ 
respectively. The negative norm states are necessarily ``more excited'', 
\ie\ have a larger contribution of the negative norm excited states 
to the total norm.

For odd number $N$ of ghost-spins, the ground state continues to be of 
the above form (\ref{Ngndst}): however it is no longer uniformly positive 
norm, since $\langle \psi_g|\psi_g\rangle$ is negative for 
$|\psi_g^+\rangle=0$.      %= |\psi_g^+|^2 + (-1)^N|\psi_g^-|^2$ 
Thus no interpretation in terms of negative norm states being ``more
excited'' is possible for $N$ odd since the ground states themselves
are not uniformly positive norm.  Instead, at each energy level, there
are equal numbers of positive and negative norm states for $N$ odd.
%\emph{Index?}

For 4 ghost-spins, the reduced density matrix for the subsystem $A$
comprising a single ghost-spin (say the first index) after tracing
over the 3 ghost-spins, generalizing (\ref{2gsrdm}), is
\be\label{rhoA4gs-1}
(\rho_A)^{\al\beta} = \gamma_{\sigma_1\rho_1} \gamma_{\sigma_2\rho_2} 
\gamma_{\sigma_3\rho_3} 
\psi^{\al\sigma_1\sigma_2\sigma_3} (\psi^*)^{\beta\rho_1\rho_2\rho_3} = 
\gamma_{\sigma_1\sigma_1} \gamma_{\sigma_2\sigma_2} \gamma_{\sigma_3\sigma_3}
\psi^{\al\sigma_1\sigma_2\sigma_3} (\psi^*)^{\beta\sigma_1\sigma_2\sigma_3}\ .
\ee
Explicitly, for the states (\ref{4gs+++-}), this becomes 
\bea\label{rdm4gs}
(\rho_A)^{++} = |\psi^{++++}|^2 - |\psi^{+++-}|^2\ , & \quad &
(\rho_A)^{+-} = 0\ , \nonumber\\
(\rho_A)^{-+} = 0\ , & \quad &
(\rho_A)^{--} = |\psi^{-+++}|^2 - |\psi^{----}|^2\ .
\eea
(We have chosen a family of states that lead to a diagonal RDM for 
convenience.)\ 
The mixed index reduced density matrix then becomes
\be
(\rho_A)^+_+ = |\psi^{++++}|^2 - |\psi^{+++-}|^2\ ,\qquad
(\rho_A)^-_- = -|\psi^{-+++}|^2 + |\psi^{----}|^2\ .
\ee
The entanglement patterns here can be analysed using the norm and setting
\be\label{rhoA4-x}
\begin{split}
  |\psi^{++++}|^2 - |\psi^{+++-}|^2 \equiv x ,\qquad 
\langle\psi|\psi\rangle = x + (\pm 1-x)\ ; \\
(\rho_A)^+_+ =  x ,\qquad (\rho_A)^-_- = \pm 1 - x ,\quad \\
\qquad\qquad S_A = - x\log x - (\pm 1 - x) \log (\pm 1 - x)\ .
\end{split}
\ee
This is very similar to the case of one spin and two ghost-spins 
analysed in \cite{Jatkar:2016lzq}, sec.~5. As in that case, we see 
that for $x>0$ we have $1-x>0$ for positive norm states, implying 
$0<x<1$, and $\rho_A$ is positive definite, with $S_A>0$.\\
If $x<0$, then $(1-x)>0$, giving 
$S_A = |x|\log |x| - (1+|x|)\log(1+|x|) + i\pi |x|$, with $Re S_A<0$ 
and $Im S_A\neq 0$ for positive norm states.\ In this case, we have 
$x<0$ implying that\ $|\psi^{++++}|^2 < |\psi^{+++-}|^2$,\ \ie\ there is 
a larger contribution from the negative norm excited state component 
$|+++-\rangle$, than the positive norm ground state $|++++\rangle$ 
component: it is these components that arise since the reduced density 
matrix involves these particular segregations of the full state.
Note that there are several fully positive norm states comprising 
other states at the first level, \eg\ $|++--\rangle,\ |--++\rangle$.
These have a positive reduced density matrix and positive entanglement.

This sort of structure is also true more generally: \eg\ for two 
ghost-spins, the generic state with norm is
\bea
&& |\psi\rangle = \psi^{++}|++\rangle + \psi^{+-}|+-\rangle + 
\psi^{-+}|-+\rangle + \psi^{--}|--\rangle\ ,\nonumber\\
&& \langle \psi|\psi\rangle = |\psi^{++}|^2 + |\psi^{--}|^2 
- |\psi^{+-}|^2 - |\psi^{-+}|^2 = \pm 1\ .
\eea
From the entanglement patterns in \cite{Narayan:2016xwq}, 
\cite{Jatkar:2016lzq}, reviewed in sec.~\ref{sec:revsGs}, we have seen 
that the subfamily (\ref{Ex:2gsEE}), (\ref{Ex:2gsEE2}), with a diagonal 
RDM shows in fact that positive norm states lead to a positive RDM and 
positive entanglement, while negative norm states have a negative 
definite RDM (all eigenvalues negative) giving $Re(EE)<0$ and 
$Im(EE)\neq 0$.
To explore the interpretation a little further, consider setting 
$\psi^{-+}=0$, \ie\ the state and reduced density matrix from 
(\ref{2gsrdm}), (\ref{rhoAc1234}), are\
\be\label{rdm++--+-}
|\psi\rangle = \psi^{++}|++\rangle + \psi^{--}|--\rangle + 
\psi^{+-}|+-\rangle\ ,\qquad
(\rho_A)^+_+ = |\psi^{++}|^2 - |\psi^{+-}|^2 ,\quad
(\rho_A)^-_- = |\psi^{--}|^2 > 0\ ,
\ee
where we suppress writing the off-diagonal components of $\rho_A$ given 
the considerations below.
For small $\psi^{+-}$, with $|\psi^{+-}|^2<|\psi^{++}|^2$, this state is 
positive norm and thus gives a positive RDM and entanglement entropy. 
However for larger $\psi^{+-}$, the effective probability $\rho^+_+$ 
decreases due to the negative norm nature of $\psi^{+-}$: for 
$|\psi^{+-}|=|\psi^{++}|$, this probability $\rho^+_+$ vanishes. Beyond 
this, the effective ``probability'' $\rho^+_+$ is negative and so the 
entanglement entropy is not positive definite.
To give further perspective, consider an observable 
$O_{s_1}=O_{s_1}^{\al\beta} |\al_1\rangle\langle \beta_1| 
= O_{s_1}^{++}|+\ran\lan +| + O_{s_1}^{--}|-\ran\lan -|$, of the first 
ghost-spin alone (for simplicity, we have taken $O_{s_1}$ to be 
diagonal with $O_{s_1}^{+-}=O_{s_1}^{-+}=0$).
Then the correlation function of $O_{s_1}$ in the state $|\psi\rangle$ is
\bea
&& \langle\psi| O_{s_1} |\psi\rangle = (\psi^*)^{\sigma\rho} 
\langle \sigma\rho| O_s^{\al_1\beta_1} |\al_1\rangle\langle \beta_1| 
\psi^{\al\beta} |\al\beta\rangle 
= (\psi^*)^{\sigma\rho} O_{s_1}^{\al_1\beta_1} \psi^{\al\beta} \
\langle\sigma|\al_1\rangle\ \langle\beta_1|\al\rangle\ \langle\rho|\beta\rangle 
\nonumber\\
&& \qquad\qquad = \gamma_{\sigma\al_1}\gamma_{\beta_1\al} \rho_A^{\sigma\al} 
O_{s_1}^{\al_1\beta_1} = (\rho_A)^\al_\beta (O_{s_1})_\al^\beta = 
(\rho_A)^+_+ (O_{s_1})_+^+ + (\rho_A)^-_- (O_{s_1})_-^-\ .
\eea
The expectation value becomes
\be
{\langle\psi| O_{s_1} |\psi\rangle\over \langle\psi |\psi\rangle}
= {(\rho_A)^+_+ (O_{s_1})_+^+ + (\rho_A)^-_- (O_{s_1})_-^-\over
(\rho_A)^+_+ + (\rho_A)^-_-}
\ee
For ordinary spins, $(\rho_A)^+_+$ and $(\rho_A)^-_-$ are always positive 
with $tr\rho_A=(\rho_A)^+_+ + (\rho_A)^-_-=1$. For ghost-spins, 
$(\rho_A)^+_+,\ (\rho_A)^-_-$ can become negative, with 
$tr\rho_A=(\rho_A)^+_+ + (\rho_A)^-_-=\pm 1$ for positive/negative norm 
states. In particular, for (\ref{rdm++--+-}) with positive norm states, 
as $(\rho_A)^+_+\ra 0$ through positive values, we have 
$(\rho_A)^-_-\ra 1$, and the expectation value becomes
\be
(\rho_A)^+_+ (O_{s_1})_+^+ + (\rho_A)^-_- (O_{s_1})_-^- 
\ \ \ \xrightarrow{\ (\rho_A)^+_+\ra 0\ }\ \ \  (O_{s_1})_-^-\ .
\ee
In other words, in the limit $|\psi^{+-}|\ra |\psi^{++}|$ with 
$(\rho_A)^+_+\ra 0$, the state behaves as if the expectation value of 
an observable cares only about the $O^-_-$ component: the $O^+_+$ 
component, in general nonzero, does not contribute. Similar phenomena 
occur with $N$ ghost-spins as well.

For $N$ ghost-spins with $N$ even, the structure of ground states and 
excited states at the first level (\ref{NgsGndSt}) (\ref{Ngs1stExcSt}) 
again suggests considering states of the form (for conveniently obtaining 
a diagonal reduced density matrix)
\be
|\psi\rangle = \psi^{+^N}|+^N\rangle + \psi^{-^N}|-^N\rangle + 
\psi^{++\ldots +-}|++\ldots +-\rangle + \psi^{--\ldots -+} |--\ldots -+\rangle
\ee
\be
|\psi^{+^N}|^2 + |\psi^{-^N}|^2 - |\psi^{++\ldots +-}|^2 
- |\psi^{--\ldots -+}|^2 = \pm 1\ .
\ee
Again the reduced density matrix for the first index ghost-spin, after 
tracing over the remaining $N-1$ ghost-spins, can be seen to have the 
simple diagonal form
\bea\label{rdmNgs}
(\rho_A)^{++} = |\psi^{+^N}|^2 - |\psi^{++\ldots +-}|^2\ , & \quad &
(\rho_A)^{+-} = 0\ , \nonumber\\
(\rho_A)^{-+} = 0\ , & \quad &
(\rho_A)^{--} = |\psi^{-+\ldots ++}|^2 - |\psi^{-^N}|^2\ ,
\eea
and the mixed index reduced density matrix becomes
\be
(\rho_A)^+_+ = |\psi^{+^N}|^2 - |\psi^{++\ldots +-}|^2\ ,\qquad
(\rho_A)^-_- = -|\psi^{-+\ldots ++}|^2 + |\psi^{-^N}|^2\ .
\ee
With $x\equiv |\psi^{+^N}|^2 - |\psi^{++\ldots +-}|^2$, the entanglement 
patterns are similar to the 4 ghost-spin case (\ref{rhoA4-x}).
Thus $x<0$ leads to $Re\, S_A<0$ and $Im\, S_A\neq 0$: $x<0$ means
$|\psi^{+^N}|^2 < |\psi^{++\ldots +-}|^2$, \ie\ there is a larger 
contribution from the negative norm excited state component in the 
RDM element.
Here also, there are several fully positive norm states comprising
other states at the first level, \eg\ $|++\ldots ++--\rangle,\
|--++\ldots ++\rangle$, with a positive reduced density matrix and
positive entanglement.

For ensembles of ghost-spins and spins, possible Hamiltonians might be
of the form\
$H = H_s + H_{gs} = -J^s\sum_{nn} s_ss_s' - J^{gs}\sum_{nn} s_{gs}s_{gs}'$,\
with the spin and ghost-spin sectors having nearest neighbour interactions
within themselves but with the spins being decoupled from the ghost-spins.
Then the ground states might be expected to be disentangled product states.

\subsection{The reduced density matrix and its eigenvalues}

Let us now study the reduced density matrix obtained after tracing
over ghost-spins and its eigenvalues in some generality.
First, it is useful to study explicit examples (\eg\ 2, 4 ghost-spins etc). 
As a simple case, consider a 2-spin state 
\be\label{psi2gs}
|\psi\rangle=c_1|++\rangle + c_2|+-\rangle + c_3|-+\rangle + c_4|--\rangle\ .
\ee
The reduced density matrix and eigenvalue equation are
\be\label{detrho2sgs}
\left( \bA{cc} c_1c_1^*\pm c_2c_2^* & \ c_1c_3^*\pm c_2c_4^* \\ 
c_3c_1^*\pm c_4c_2^* & \ c_3c_3^*\pm c_4c_4^* \eA\right) 
\quad\longrightarrow\quad
{\rm det} \left( \bA{cc} c_1c_1^*\pm c_2c_2^* -\lambda & c_1c_3^*\pm c_2c_4^* 
\\ c_3c_1^*\pm c_4c_2^* & c_3c_3^*\pm c_4c_4^* \mp \lambda \eA\right) = 0
\ee
where the top signs correspond to ordinary spins, while the bottom
signs pertain to ghost-spins. For ordinary spins, the basis states are
all positive norm while for ghost-spins, a single minus sign gives
negative norm as we have seen.
Thus for an ensemble of ordinary spins, the eigenvalue equation is\ 
\be
{\rm det} \big( (\rho_A)^{jk} - \lambda \delta^{jk} \big) = 0 \quad 
\longrightarrow\quad
\lambda^2 - ({\rm tr}\rho_A) \lambda + {\rm det}\rho_A = 0\ .
\ee
Since there are only positive norm states here, $tr\rho_A=1$ giving
\be
\lambda^2 - \lambda - {\rm det}\rho_A = 
\Big(\lambda - {1\over 2}\Big)^2 - {1\over 4} + {\rm det}\rho_A = 0\quad
\Rightarrow\quad \lambda - {1\over 2} 
= \pm \sqrt{{1\over 4} - {\rm det}\rho_A}\ .
\ee
Simplifying from (\ref{detrho2sgs}) above, it can be seen that
\be
{\rm det}\rho_A = |c_1c_4-c_2c_3|^2\ .
\ee
Since the norm condition on the state for ordinary spins is\ 
$\sum_i|c_i|^2=1$, we see that each $c_i$ is bounded with\ 
$0\leq |c_i|\leq 1$ implying that\ ${\rm det}\rho_A$ is bounded (with 
maximum value ${1\over 4}$). This in turn implies that the eigenvalues 
$\lambda$ are always real.

Now consider ghost-spins: the eigenvalue equation for the mixed index 
reduced density matrix is
\be
(\rho_A)_i^k e_k = \lambda e_i = \lambda \delta_i^ke_k\qquad\Rightarrow\qquad 
(\rho_A)^{ij} e_j = \gamma^{ik}(\rho_A)_k^j e_j = \gamma^{ij} \lambda e_j\ ,
\ee
giving 
\be\label{eveqngs}
{\rm det} \big( (\rho_A)^{jk} - \lambda \gamma^{jk} \big) = 0 \quad 
\longrightarrow\quad
\lambda^2 - ({\rm tr}\rho_A) \lambda - {\rm det}\rho_A^{ik} = 0\ .
\ee
Since $tr\rho_A=(\rho_A)^{++}-(\rho_A)^{--}=\pm 1$ for positive/negative 
norm states respectively, this becomes
\bea\label{evgs}
\lambda^2 \mp \lambda - {\rm det}\rho_A^{ik} = 0 
&=& \Big(\lambda \mp {1\over 2}\Big)^2 - {1\over 4} - {\rm det}\rho_A^{ik}
\qquad\Rightarrow \nonumber\\
\lambda =  {1\over 2} \pm \sqrt{{1\over 4} + {\rm det}\rho_A^{ik}}\ \ \
[+ve\ norm] ; &&
\lambda =  -{1\over 2} \pm \sqrt{{1\over 4} + {\rm det}\rho_A^{ik}}\ \ \
[-ve\ norm]\ .\quad 
\eea
From (\ref{detrho2sgs}) specializing to 2 ghost-spins, it can be seen
that
\be
{\rm det}\rho_A^{ik} = -|c_1c_4-c_2c_3|^2\ .
\ee
However in this case, the norm condition gives
\be
|c_1|^2 + |c_4|^2 - |c_2|^2 - |c_3|^2 = \pm 1\ ,
\ee
so that $|c_i|$ are not forced to be bounded, but in fact can be 
arbitrarily large while retaining the norm condition. For positive 
norm states, we have $|c_1|, |c_4| > |c_2|, |c_3|$, with $0\leq |c_i|<\infty$, 
while for negative norm states, we have $|c_2|, |c_3| > |c_1|, |c_4|$.
This makes the determinant potentially unbounded, as we will see below. 
From the norm condition, we see that for positive norm states, the $c_i$ 
can be parametrized as\ 
$c_1=\cosh\theta\cos\phi_1 e^{i\al_1},\ c_4=\cosh\theta\sin\phi_1 e^{i\al_4},\
c_2=\sinh\theta\cos\phi_2 e^{i\al_2},\ c_3=\sinh\theta\sin\phi_2 e^{i\al_3}$,\
while for negative norm states, the parametrizations can be switched 
as $c_1, c_4 \leftrightarrow c_2,c_3$. On the real slice, we have all 
$c_i$ real, \ie\ the phases $\al_i$ are all zero.

The minus signs in the norm make the determinant behaviour non-uniform: 
there are several branches. It is easiest to illustrate this on the real 
slice with all $c_i$ real. First consider the 1-parameter family of
states (\ref{Ex:2gsEE}) which give a diagonal reduced density matrix: 
from (\ref{detrho2sgs}), these have
\bea
\rho_A^{+-}=0=c_1c_3-c_2c_4 \quad \Rightarrow\quad
(c_1^2-c_2^2) \Big(1+{c_4^2\over c_1^2}\Big) = \pm 1\ ,\nonumber\\
{\rm and}\quad \rho^{++}=\pm x\ ,\quad \rho^{--}=\mp (1-x)\ ,\qquad 
x={c_1^2\over c_1^2+c_4^2}\ ,
\eea
and
\be
{\rm det}\rho_A^{ik} = -\Big|c_1c_4 - c_2\Big({c_2c_4\over c_1}\Big)\Big|^2 
= - {c_1^2c_4^2\over (c_1^2+c_4^2)^2} = -x(1-x)\ .
\ee
It can be seen now that
\be
{1\over 4} + {\rm det}\rho_A^{ik} = {1\over 4}-x(1-x) 
= \Big( x-{1\over 2}\Big)^2 > 0\ ,
\ee
so that the eigenvalues $\lambda$ on this diagonal 1-parameter branch 
are always real. At the point $x={1\over 2}$ the eigenvalues are both 
${1\over 2}$ for positive norm states.

On the other hand, a distinct branch arises taking $c_3=0$, \ie\ the states 
(\ref{rdm++--+-}): this gives
\be
c_1^2+c_4^2-c_2^2=\pm 1\ ,\qquad {\rm det}\rho_A^{ik}=-|c_1c_4|^2\ .
\ee
For $c_2=0$, this state is positive norm: it continues to be positive 
norm for small nonzero $c_2$ so $det\rho_A^{ik}$ is small as well and the 
eigenvalues continue to be real. However for $c_1, c_4, c_2$ large 
satisfying the norm condition, the determinant is large and negative 
rendering the eigenvalues $\lambda$ complex using (\ref{evgs}), even for 
positive norm states. The real and complex $\lambda$ branches intersect 
at a locus of coinciding eigenvalues $\lambda={1\over 2}$ when 
$det\rho_A^{ik}=-{1\over 4}$.

Thus we see that for 2 ghost-spins, the reduced density matrix exhibits 
several distinct branches with the eigenvalue spectrum varying from 
real to complex, much unlike the ordinary spin case.

It can also be checked that for an ensemble of one spin and two
ghost-spins, tracing over all the ghost-spins leads to a reduced
density matrix for the spin alone whose eigenvalue equation is again
of the form above, for spins alone.

\noindent {\bf Zero norm states}:\ \ 
There are zero norm states in ghost-spin systems of the sort we have been 
discussing: \eg\ states $|\ua\ran$ and $|\da\ran$ are zero norm themselves.
To study entanglement in these cases, consider the 2 ghost-spin case: 
zero norm states (\ref{psi2gs}) have 
$\langle\psi|\psi\rangle=0$, \ie\
\be
|c_1|^2+|c_4|^2=|c_2|^2+|c_3|^2\  \qquad {\rm and}\qquad  tr\rho_A=0\ .
\ee
We also have\ $det\rho_A^{ik}=-|c_1c_4-c_2c_3|^2$ as above. So the 
eigenvalue equation (\ref{eveqngs}) is\ 
\be
\lambda^2=det\rho_A^{ik} <0
\ee
since $tr\rho_A=0$ and the eigenvalues are always pure imaginary. 
The entanglement entropy can of course have real and imaginary parts
on evaluating this. Again $det\rho_A^{ik}$ can acquire large negative
values: \eg\ on the branch $c_3=0$, we have $det\rho_A^{ik}=-|c_1c_4|^2$
which becomes large and negative when $c_1,c_4$ are large.
Most basically however, zero norm states do not have any canonical 
normalization: an overall scaling changes the $c_i$ and therefore 
$\lambda$ as well.

\subsection{RDM, eigenvalues and $\ua\leftrightarrow\da$ symmetry}

In the $\ua, \da$-basis, we have $\gamma_{\ua\da}=\gamma_{\da\ua}=1$: 
then the 2 ghost-spin state (\ref{psi2gs}) is
\be
|\psi\ran = \psi^{\da\da}|\da\da\ran + \psi^{\ua\ua}|\ua\ua\ran + 
\psi^{\da\ua}|\da\ua\ran + \psi^{\ua\da}|\ua\da\ran
\ee
with norm 
\be
\lan\psi|\psi\ran = (\psi^*)^{\ua\ua}\psi^{\da\da} 
+ (\psi^*)^{\da\da}\psi^{\ua\ua}
+ (\psi^*)^{\ua\da}\psi^{\da\ua} 
+ (\psi^*)^{\da\ua}\psi^{\ua\da} = \pm 1\ .
\ee
The reduced density matrix 
$(\rho_A)^{\alpha\kappa}=\gamma_{\beta\lambda}\psi^{\alpha\beta}
{\psi^{\kappa\lambda}}^*$ after tracing over the second ghost-spin is
\bea\label{detrho2sgs-ud}
&& 
\left( \bA{cc} (\psi^*)^{\ua\ua}\psi^{\ua\da} + (\psi^*)^{\ua\da}\psi^{\ua\ua} 
& \ (\psi^*)^{\ua\ua}\psi^{\da\da} + (\psi^*)^{\ua\da}\psi^{\da\ua} \\ 
(\psi^*)^{\da\ua}\psi^{\ua\da} + (\psi^*)^{\da\da}\psi^{\ua\ua}
& \ (\psi^*)^{\da\ua}\psi^{\da\da} + (\psi^*)^{\da\da}\psi^{\da\ua}
\eA\right) 
\quad\longrightarrow\quad \nonumber\\
&& {\rm det} \left( 
\bA{cc} (\psi^*)^{\ua\ua}\psi^{\ua\da} + (\psi^*)^{\ua\da}\psi^{\ua\ua} 
& \ (\psi^*)^{\ua\ua}\psi^{\da\da} + (\psi^*)^{\ua\da}\psi^{\da\ua} - \lambda\\ 
(\psi^*)^{\da\ua}\psi^{\ua\da} + (\psi^*)^{\da\da}\psi^{\ua\ua} - \lambda
& \ (\psi^*)^{\da\ua}\psi^{\da\da} + (\psi^*)^{\da\da}\psi^{\da\ua}
\eA\right)  = 0\ .
\eea
The eigenvalue equation in the second line becomes
\be
\lambda^2-({\rm tr}\rho_A^{ik})\lambda - {\rm det}\rho_A^{ik} = 0
\ee
with 
\be
tr\rho_A=\rho_A^{\ua\da}+\rho_A^{\da\ua}=\lan\psi|\psi\ran = \pm 1\ ,
\qquad {\rm det}\rho_A^{ik} 
= -\big|\psi^{\da\da}\psi^{\ua\ua}-\psi^{\da\ua}\psi^{\ua\da}\big|^2\ .
\ee
So for a state ${1\over\sqrt{2}}(|\ua\ua\ran\pm|\da\da\ran)$, we obtain 
$\rho_A=(^0_{\pm 1/2}\ {}^{\pm1/2}_0)$ giving\ $(\lambda\mp {1\over 2})^2=0$
which is $\lambda={1\over 2} , {1\over 2} $ for $+ve$ norm and 
$\lambda=-{1\over 2} , -{1\over 2} $ for $-ve$ norm. Other previous 
cases can be recast in this basis as well.

Consider now a symmetry $\ua\leftrightarrow\da$ which exchanges up and 
down ghost-spins. Retaining only states invariant under this 
$\ua\leftrightarrow\da$ symmetry, the general state above collapses to
\be
|\psi\ran = \psi^{\da\da}\left(|\da\da\ran + |\ua\ua\ran\right) + 
\psi^{\da\ua}\left( |\da\ua\ran + |\ua\da\ran \right)
\ee
and the norm is
\be
\lan\psi|\psi\ran = 2 |\psi^{\da\da}|^2 + 2 |\psi^{\da\ua}|^2 > 0\ ,\qquad
{\rm tr}\rho_A^{ik}=\lan\psi|\psi\ran=+1\ .
\ee
This is always positive definite. The RDM above becomes
\be
\rho_A^{ik} = 
\left( \bA{cc} (\psi^*)^{\da\ua}\psi^{\da\da} + (\psi^*)^{\da\da}\psi^{\da\ua}
& \  |\psi^{\da\da}|^2 + |\psi^{\da\ua}|^2 \\ 
|\psi^{\da\da}|^2 + |\psi^{\da\ua}|^2
& \ (\psi^*)^{\da\ua}\psi^{\da\da} + (\psi^*)^{\da\da}\psi^{\da\ua}
\eA\right)
\ee
The determinant is
\be
{\rm det}\rho_A^{ik} 
= -\Big|{1\over 2} - 2(\psi^{\da\da})^2 \Big|^2
\ee
so
\be
\lambda^2-\lambda + \Big|{1\over 2} - 2(\psi^{\da\da})^2 \Big|^2 = 0\ \quad
\Rightarrow\quad
\Big(\lambda - {1\over 2}\Big)^2 
= {1\over 4} - \Big|{1\over 2} - 2(\psi^{\da\da})^2 \Big|^2 > 0\ .
\ee
This is quite like the case for ordinary spins. Since 
$|\psi^{\da\da}|<{1\over 2}$, we have the determinant bounded and so 
$\lambda$ above is real, positive and bounded with 
$\sum_i\lambda_i=1$. So $S_A>0$.

In the $\pm$-basis, the $\ua\leftrightarrow\da$ symmetry is even 
more strikingly simple: we see that the $|-\ran$ state simply collapses 
as $|-\ran = {1\over\sqrt{2}}(|\ua\ran-|\da\ran)\ra 0$ 
leaving only the $|+\ran$ state which is positive definite. Thus truncating 
all states in any ensemble of ghost-spins to only those invariant under 
$\ua\leftrightarrow\da$ symmetry renders the ghost-spin Hilbert space 
manifestly positive definite.

\subsection{Modified inner product and unitarity}

It is known that various non-Hermitian Hamiltonians admit $PT$-symmetric 
extensions \cite{Bender:1998ke,Bender:2007nj} which render the system
unitary. In light of the fairly ordinary looking positive definite
ghost-spin partition functions \eg\ (\ref{2gsZ2}), (\ref{NgsZ}), it is
interesting to ask if there is a modified inner product that leads to
an effectively unitary structure for these systems (see \eg\ 
\cite{LeClair:2007iy} for similar discussions in the context of 3-dim 
symplectic fermion theories).
Consider introducing an operator $C$ such that nonzero expectation 
values are obtained only after a $C$ insertion: \ie\
\be
\langle\downarrow |\downarrow\rangle = 0 = \langle\uparrow |\uparrow\rangle ,
\qquad \langle \downarrow |C | \downarrow\rangle = 1 = 
\langle\uparrow|C|\uparrow\rangle\ .
\ee
Then a generic ghost-spin state and its adjoint can be defined as
\be
|\psi\rangle = c_1 |\uparrow\rangle + c_2 |\downarrow\rangle\ ,\qquad
\left( |\psi\rangle \right)^\dag = c_1^* \langle \uparrow|
+ c_2^* \langle \downarrow|\ .
\ee
Using the above inner products with $C$ insertions, we have the 
modified inner product for the state as
\be
\left( (|\psi\rangle)^\dag, |\psi\rangle \right) \equiv 
\langle \psi|C| \psi \rangle = c_1 c_1^* \langle\uparrow|C|\uparrow\rangle 
+ c_2 c_2^* \langle \downarrow |C | \downarrow\rangle = |c_1|^2+|c_2|^2\ ,
\ee
which is positive definite, thus defining a unitary structure on the 
Hilbert space.
Thus all states are now positive norm: in particular the $|\pm\rangle$ 
states have norm
\be
\langle\pm|C|\pm\rangle\ \propto\ \langle\uparrow|C|\uparrow\rangle 
+ \langle \downarrow |C| \downarrow\rangle > 0\ .
\ee
The fact that the partition functions previously discussed resemble
those for an ordinary spin system can be taken to imply the existence
of such an operator $C$ and the above unitary modification of the
inner product to be positive definite. With this unitary inner product, 
the reduced density matrix for any subsystem of ghost-spin states is 
always positive definite and therefore so is the entanglement entropy.

Now consider coupling an ensemble of ghost-spins to an ensemble of 
ordinary spins. Define the inner product on states to be the usual one 
for the spin sector and to be the above unitary inner product on the 
ghost-spin sector. For instance in the one spin and two ghost-spins 
system, a family of states (which formerly contained negative norm states) 
and the associated inner product then are
\bea
&& |\psi\rangle = \psi^{\uparrow,++}|\uparrow,++\rangle 
+ \psi^{\uparrow,+-}|\uparrow,+-\rangle 
+ \psi^{\downarrow,-+}|\downarrow,-+\rangle 
+ \psi^{\downarrow,--}|\downarrow,--\rangle\ ,\nonumber\\
&& \langle\psi|\psi\rangle = \ldots + |\psi^{\uparrow,+-}|^2 
+ |\psi^{\downarrow,-+}|^2 > 0\ ,
\eea
using the above unitary inner products for $|\pm\rangle$. This is 
always positive definite so there are no negative norm states. 
In fact this now maps the spin and 2 ghost-spins system to a system of 
3 ordinary spins.  However the physical system originally was a single
spin coupled with 2 ghost-spins: the ghost-spins are regarded as
unphysical, reflecting the negative norm subsector arising from fixing
a gauge symmetry. The physical subsector therefore is the single
spin. From this point of view, the mapping to a system of 3 ordinary
spins is a formal process since the physical subspace of the original
system continues to be the single spin. The modified inner product in
the ``ghost spin'' sector unitarises the system.  This process in our
case turns out to be a formal tool to ``explain'' why we get
relatively ordinary looking partition functions for our choice of the
Hamiltonian. So we will not pursue this $PT$-symmetric formulation 
further here.

Interesting generalizations of the finite ghost-spin chains we have
been studying so far involve infinite ghost-spin chains and their
possible continuum limits at criticality where a conformal field
theory may emerge. We will study one concrete class of examples in the
next section.

\section{Ghost-spin chains and the $bc$-ghost CFT}

In this section we will look at a family of infinite ghost-spin chains
with a different interaction, although still based on the ghost-spins
used so far treated as the underlying microscopic variables. Motivated
by the well-known fact that the Ising spin chain at criticality is
described by a CFT of free massless fermions (see \eg\
\cite{Kogut:1979wt},\cite{SachdevQPT}), one might expect that infinite
ghost-spin chains exhibit critical points at which a continuum
description of the chain exists in terms of ghost-CFTs such as the
$bc$-CFT (discussed extensively in \eg\
\cite{polchinskiTextBk,Blumenhagen:2013fgp}, as well as
\cite{Friedan:1985ge}, and more recently \cite{Saleur:1991hk,
  Kausch:1995py,Kausch:2000fu,Flohr:2001zs,Krohn:2002gh}). The
off-diagonal inner products for states here reflect the off-diagonal
oscillator algebra $\{b_n, c_m\} = \delta_{n+m,0}$ of the $bc$-ghost
CFT.

In this light, consider an infinite 1-dimensional ghost-spin chain
with a nearest neighbour interaction Hamiltonian
\be\label{Hgssigmabc0}
H = J \sum_n \left( \sigma_{b(n)} \sigma_{c(n+1)} +
  \sigma_{b(n)}\sigma_{c(n-1)} \right)\ , 
\ee 
where $\sigma_{bn}$ and $\sigma_{cn}$ are two species of 2-state spin
variables defined at each site and $n$ labels the lattice site in the
chain.  The nearest neighbour interaction in this Hamiltonian is more
akin to a hopping type interaction than the Ising type Hamiltonian in
(\ref{2gsIsingHint}), (\ref{NgsIsingHint}): we will describe this in
detail later.  The spin variables $\sigma_{bn},\ \sigma_{cn}$ satisfy
the (anti-)commutation relations
\be\label{sigmaCommRelns}
\{ \sigma_{bn},\ \sigma_{cn} \} = 1\ ,\qquad [\sigma_{bn}, \sigma_{bn'}] =
[\sigma_{cn}, \sigma_{cn'}] = [\sigma_{bn}, \sigma_{cn'}] = 0\ ,
\ee
which are consistent with the off-diagonal inner product between
ghost-spin states.  These spin variables are self-adjoint and act on the 
two states $|\ua\ran,\ |\da\ran$, at each lattice site $n$, as 
\be
\sigma_{bn}^\dag=\sigma_{bn} ,\ \ 
\sigma_{cn}^\dag=\sigma_{cn}\ ;\qquad \sigma_b|\da\ran = 0 ,\quad
\sigma_b|\ua\ran = |\da\ran ,\quad \sigma_c|\ua\ran = 0 ,\quad
\sigma_c|\da\ran = |\ua\ran .  
\ee 
Thus the $\sigma_{bn}$ act as
lowering operators while the $\sigma_{cn}$ act as raising
operators. It is worth noting that the $\sigma_b, \sigma_c$ cannot be
$2\times 2$ Pauli matrices, since the latter satisfy
$\{ \sigma^-,\ \sigma^+ \} = 1$ but with $(\sigma^-)^\dag = \sigma^+$.
The present algebra is off-diagonal, with hermitian operators.
%% \emph{[NOTE:\ $4\times 4$ matrix representations?]}

As an example, for 2 ghost-spins, we have 4 states, 
$|\da\da\ran,\ |\da\ua\ran,\ |\ua\da\ran,\ |\ua\ua\ran$. 
These states can be expressed as
\be
|\da\ua\ran = \sigma_{c2}|\da\da\ran ,\quad 
|\ua\da\ran = \sigma_{c1}|\da\da\ran ,\qquad
|\ua\ua\ran = \sigma_{c1}\sigma_{c2}|\da\da\ran 
= \sigma_{c2}\sigma_{c1}|\da\da\ran ,
\ee
the last expression implying that the $\ua$-excitations in $|\ua\ua\ran$ 
have no particular order. As is natural for spin systems, the 
spin $\sigma$-variables at distinct lattice sites commute as in 
(\ref{sigmaCommRelns}), \eg\ $\sigma_{cn}\sigma_{cn'}=\sigma_{cn'}\sigma_{cn}$.
Now with the off-diagonal inner-products between states, we have
\be\label{2gssigmaNorms}
\lan\ua\ua|\da\da\ran = 1 = \lan\da\da|\ua\ua\ran\ ,\qquad
\lan\da\ua|\da\ua\ran = 1 = \lan\ua\da|\ua\da\ran\ .
\ee
In the second set of inner products, note that the spins have been 
ordered right to left in the bra states: this is distinct from that 
used throughout the paper so far, \eg\ (\ref{2gsnorms}). We have 
re-ordered in this manner anticipating our description of fermionic 
excitations in what follows.

The inner products above can be written explicitly in terms of the spin 
operators as \eg\
\be
\lan\da\da|\ua\ua\ran = \lan\da\da| \sigma_{c1}\sigma_{c2}|\da\da\ran = 1\ ,
\qquad \lan\ua\da|\ua\da\ran 
= \lan\da\da| \sigma_{c2}^\dag \sigma_{c1}|\da\da\ran = 1\ ,
\ee
and so on, using $\sigma_{ci}^\dag=\sigma_{ci}$:\ in this form, the 
ordering of spin operators is unimportant since they are commuting,
however, it will be relevant once we have fermionic representations of
these states.  Now a basis of positive and negative norm states for 
2 ghost-spins is
\bea\label{2gsbasis:e12+-}
&& |e^1_\pm\ran={1\over\sqrt{2}}|\da\da\ran\pm|\ua\ua\ran\quad\longrightarrow
\quad  \lan e^1_\pm|e^1_\pm\ran = \pm {1\over 2} ( \lan\ua\ua|\da\da\ran + 
\lan\da\da|\ua\ua\ran ) = \pm 1\ ,\nonumber\\
&& |e^2_\pm\ran={1\over\sqrt{2}}|\da\ua\ran\pm|\ua\da\ran\quad\longrightarrow
\quad  \lan e^2_\pm|e^2_\pm\ran = \pm {1\over 2} ( \lan\da\ua|\da\ua\ran + 
\lan\ua\da|\ua\da\ran ) = \pm 1\ .
\eea

%\vspace{5mm}

\subsection{Ghost-spin chains and fermionic excitations}

We want to construct fermionic operators out of the commuting spin 
operators $\sigma_b$, $\sigma_c$.  This can be achieved using a version 
of the Jordan-Wigner transformation \cite{Kogut:1979wt,SachdevQPT}, 
which we will describe in the next subsection.
These fermionic operators satisfy anti-commutation relations
\be\label{fermgsops-abc}
\{ a_{bi}, a_{cj} \} = \delta_{ij}\ ,\qquad 
\{ a_{bi}, a_{bj} \} = 0\ , \qquad  \{ a_{ci}, a_{cj} \} = 0\ .
\ee
So in particular unlike the $\sigma$ spin operators, these 
anticommute not just at the same site $i$ but also at distinct sites $i, j$.\
The ket and bra states exhibit the action
\bea
&& a_b|\da\ran = 0 ,\qquad a_b|\ua\ran = |\da\ran ,\qquad
a_c|\ua\ran = 0 ,\qquad a_c|\da\ran = |\ua\ran ,\nonumber\\
&& \lan \da| a_b = 0 ,\qquad \lan \ua| a_b = \lan\da| ,\qquad
\lan \da| a_c = \lan \ua| ,\qquad \lan \ua| a_c = 0\ .
\eea
Now to construct ket states and their corresponding bra states, we have 
to be careful about the ordering of the operators and the spin excitations 
at the various sites, especially in constructing inner products of states. 
We adopt the convention that 
\bea\label{fgsOrdering}
&& \lan\ua\ua| \da\da\ran = 1\ ;\qquad\quad
|\ura{\ua\ua}\ran = a_{c1} a_{c2} |\da\da\ran;\qquad\ \
\lan\ula{\da\da}| = \lan \ua\ua| a_{b2} a_{b1}\ ,\nonumber\\
&& \Rightarrow \qquad \lan\ula{\da\da} |\ura{\ua\ua}\ran 
= \lan \ua\ua| a_{b2} a_{b1}\ a_{c1} a_{c2} |\da\da\ran 
= \lan\ua\ua| \da\da\ran = 1\ ,
\eea
where we have illustrated two fermionic ghost-spins for simplicity.
In other words, the underlining right arrow below the spins in the ket
state displays the order of the operator excitations to be increasing
to the right, and the underlining left arrow below the spins in the
bra state shows the order to be increasing to the left. The states
$|\da\da\ran$ and $\lan\ua\ua|$ above are the empty and filled ket and
bra states respectively so the ghost-spins in them are not underlined
since they do not need ordering.\
Likewise for three fermionic ghost-spins, we have
\bea
&& \lan\ua\ua\ua| \da\da\da\ran = 1\ ;\qquad\quad
|\ura{\ua\ua\ua}\ran = a_{c1} a_{c2} a_{c3} |\da\da\da\ran;\qquad\ \
\lan\ula{\da\da\da}| = \lan \ua\ua\ua| a_{b3} a_{b2} a_{b1}\ ,\nonumber\\
&& \Rightarrow \qquad \lan\ula{\da\da\da} |\ura{\ua\ua\ua}\ran 
= \lan \ua\ua\ua| a_{b3} a_{b2} a_{b1}\ a_{c1} a_{c2} a_{c3} |\da\da\da\ran 
= \lan\ua\ua\ua| \da\da\da\ran = 1\ .
\eea

The intuition here is that the ket state being $|\prod_i\da_i\ran$
corresponds to an empty state, and then an $a_{ci}$ operator acts on
it to the right to fill it with a ``particle''-like
$\ua_i$-excitation. These being fermionic have to be ordered towards
the right. By contrast, the bra state corresponding to
$\lan\prod_i\ua_i|$ is a ``filled'' state and then an $a_{bi}$ operator
acts on it to the left to remove a $\ua_i$-excitation or create a
``hole''-like $\da_i$-excitation. The $a_{bi}$ are ordered increasing
towards the left.

Let us now focus on two fermionic ghost-spins and explore further.
A state of the form below and its adjoint defined appropriately are
\bea\label{2fgsStAdj}
&& |\psi\ran = \psi_1|\da\da\ran + \psi_2 |\ura{\ua\ua}\ran = 
\psi_1 |\da\da\ran + \psi_2 a_{c1} a_{c2} |\da\da\ran\ ,\nonumber\\
&& \lan\psi| = \psi_1^* \lan\ula{\da\da}| + \psi_2^* \lan\ua\ua| = 
\psi_1^* \lan\ua\ua| a_{b2} a_{b1} + \psi_2^* \lan\ua\ua|\ .
\eea
The first expression in each line is written purely in terms of the 
ordered fermionic ghost-spin basis states while the second expression 
expresses this in terms of the fermionic ghost-spin operators ordered 
appropriately, with the spins in the bra going right to left as the 
underlining arrow indicates. The inner product of these states then is
\bea
\lan\psi| \psi\ran &=& \psi_1^*\psi_2 \lan\ula{\da\da} |\ura{\ua\ua}\ran 
+ \psi_2^*\psi_1 \lan\ua\ua |\da\da\ran \nonumber\\
&=& 
\psi_1^*\psi_2 \lan\ua\ua| a_{b_2} a_{b1}\ a_{c1} a_{c2} |\da\da\ran 
+ \psi_2^*\psi_1 \lan\ua\ua |\da\da\ran = \psi_1^*\psi_2 + \psi_2^*\psi_1\ .
\eea
This is the expected indefinite norm so the system contains negative
norm states: for instance $|\da\da\ran - |\ura{\ua\ua}\ran$ has norm
$-2$. This definition of the adjoints (\ref{2fgsStAdj}) is consistent
with the off-diagonal inner products of the commuting spin states
(\ref{2gssigmaNorms}).

The rule for constructing the adjoint state is to write the bra state
with the spins written as in the ket, but ordered right to left (along 
the underlining arrow in the bra states). The states $|\da\da\ran$ and 
$\lan\ua\ua|$ as stated below (\ref{fgsOrdering}) do not need ordering, 
while for instance, the ket $|\ura{\da\ua}\ran$ has adjoint 
$\lan\ula{\ua\da}|$. Thus using these basis states, we have states and 
their adjoints,
\bea
&& |\psi\ran = \psi_1|\ura{\da\ua}\ran + \psi_2 |\ura{\ua\da}\ran = 
\psi_1 a_{c2} |\da\da\ran + \psi_2 a_{c1} |\da\da\ran\ ,\nonumber\\
&& \lan\psi| = \psi_1^* \lan\ula{\ua\da}| + \psi_2^* \lan\ula{\da\ua}| = 
\psi_1^* \lan\ua\ua| a_{b1} + \psi_2^* \lan\ua\ua| a_{b2}\ .
\eea
The inner product is
\bea
\lan\psi| \psi\ran &=& \psi_1^*\psi_2 \lan\ula{\ua\da} |\ura{\ua\da}\ran 
+ \psi_2^*\psi_1 \lan\ula{\da\ua} |\ura{\da\ua}\ran \nonumber\\
&=& \psi_1^*\psi_2 \lan\ua\ua| a_{b1} a_{c1} |\da\da\ran +
\psi_2^*\psi_1 \lan\ua\ua| a_{b2} a_{c2} |\da\da\ran = 
\psi_1^*\psi_2 + \psi_2^*\psi_1\ .
\eea
This is again the expected indefinite norm:
\eg\ $|\ura{\da\ua}\ran - |\ura{\ua\da}\ran$ is a negative norm state 
with norm $-2$. We see that a state $|\ura{\da\ua}\ran$ (with $\psi_2=0$) 
has adjoint $\lan\ula{\ua\da}|$ and zero norm since 
$\lan\psi|\psi\ran = \lan\ula{\ua\da} |\ura{\da\ua}\ran =
\lan\ua\ua| a_{b1} a_{c2} |\da\da\ran=0$.

Consider now 3 fermionic ghost-spins, and states/adjoints,
\bea
&& |\psi\ran = \psi_1|\ura{\da\ua\da}\ran + \psi_2 |\ura{\ua\da\ua}\ran = 
\psi_1 a_{c2} |\da\da\da\ran + \psi_2 a_{c1}a_{c3} |\da\da\da\ran\ ,\nonumber\\
&& \lan\psi| = \psi_1^* \lan\ula{\da\ua\da}| + \psi_2^* \lan\ula{\ua\da\ua}| = 
\psi_1^* \lan\ua\ua\ua| a_{b3}a_{b1} + \psi_2^* \lan\ua\ua\ua| a_{b2}\ .
\eea
The norm of this state $|\psi\ran$ is given by the inner product
\bea
\lan\psi| \psi\ran &=& \psi_1^*\psi_2 \lan\ula{\da\ua\da} |\ura{\ua\da\ua}\ran 
+ \psi_2^*\psi_1 \lan\ula{\ua\da\ua} |\ura{\da\ua\da}\ran \nonumber\\
&=& \psi_1^*\psi_2 \lan\ua\ua\ua| a_{b3}a_{b1}~a_{c1}a_{c3} |\da\da\da\ran +
\psi_2^*\psi_1 \lan\ua\ua\ua| a_{b2} a_{c2} |\da\da\da\ran = 
\psi_1^*\psi_2 + \psi_2^*\psi_1\ ,\qquad
\eea
which is the expected indefinite norm.\ Along these lines, note that 
a state of the form\ 
$|\psi\ran = |\ura{\da\ua\ua}\ran=a_{c2}a_{c3}|\da\da\da\ran$\ has its 
adjoint $\lan\psi|=\lan\ula{\ua\ua\da}| = \lan\ua\ua\ua| a_{b1}$: this 
has zero norm, since\ 
$\lan\psi|\psi\ran = \lan\ula{\ua\ua\da} |\ura{\da\ua\ua}\ran=
\lan\ua\ua\ua| a_{b1} a_{c2}a_{c3} |\da\da\da\ran=0$.

Likewise for 4 fermionic ghost-spins, states/adjoints of the form
\bea
&& |\psi\ran = \psi_1|\ura{\da\ua\da\ua}\ran + \psi_2 |\ura{\ua\da\ua\da}\ran 
= \psi_1a_{c2}a_{c4}|\da\da\da\da\ran + \psi_2a_{c1}a_{c3}|\da\da\da\da\ran\ ,
\nonumber\\
&& \lan\psi| = \psi_1^* \lan\ula{\ua\da\ua\da}| + 
\psi_2^* \lan\ula{\da\ua\da\ua}| = \psi_1^* \lan\ua\ua\ua\ua| a_{b3}a_{b1} 
+ \psi_2^* \lan\ua\ua\ua\ua| a_{b4}a_{b2}\ ,
\eea
have norm given by the inner product
\be
\lan\psi| \psi\ran =
\psi_1^*\psi_2 \lan\ua\ua\ua\ua| a_{b3}a_{b1}~a_{c1}a_{c3} |\da\da\da\da\ran +
\psi_2^*\psi_1 \lan\ua\ua\ua\ua| a_{b4}a_{b2} a_{c2}a_{c4} |\da\da\da\da\ran = 
\psi_1^*\psi_2 + \psi_2^*\psi_1\ .
\ee
We see that a state $|\ura{\da\ua\da\ua}\ran$ (with $\psi_2=0$) is then 
zero norm, its adjoint being $\lan\ula{\ua\da\ua\da}|$.

\subsection{Ghost-spins and a Jordan-Wigner transformation}

As stated earlier, we want to start with the ghost-spin chain
described in terms of the commuting spin $\sigma_b,\sigma_c$-variables 
and go to the fermionic ghost-spin $a_b, a_c$-variables. Consider the
following generalization of the usual Jordan-Wigner transformation 
\cite{Kogut:1979wt,SachdevQPT}, written here for the commuting 
ghost-spin variables,
\bea\label{JWsigmabcnabcn}
&& \sigma_{b1}=a_{b1}\ ,\quad \sigma_{c1}=a_{c1}\ ,\qquad 
\sigma_{b2}=i(1-2a_{c1}a_{b1})a_{b2}\ ,\quad 
\sigma_{c2}=-i(1-2a_{c1}a_{b1})a_{c2}\ ,\ \ \ \ldots\ ,\nonumber\\
&& 
\sigma_{bn}=i(1-2a_{c1}a_{b1})i(1-2a_{c2}a_{b2})\ldots 
i(1-2a_{c(n-1)}a_{b(n-1)})a_{bn}\ ,\nonumber\\
&& 
\sigma_{cn}=(-i)(1-2a_{c1}a_{b1})(-i)(1-2a_{c2}a_{b2})\ldots 
(-i)(1-2a_{c(n-1)}a_{b(n-1)})a_{cn}\ ,\ \ \ \ldots
\eea
The inverse transformations for the fermionic ghost-spin variables are
\bea\label{JWabcnsigmabcn}
&& a_{b1}=\sigma_{b1}\ ,\quad a_{c1}=\sigma_{c1}\ ,\qquad 
a_{b2}=i(1-2\sigma_{c1}\sigma_{b1})\sigma_{b2}\ ,\quad 
a_{c2}=-i(1-2\sigma_{c1}\sigma_{b1})\sigma_{c2}\ ,\ \ \ \ldots ,\nonumber\\
&& 
a_{bn}=i(1-2\sigma_{c1}\sigma_{b1})i(1-2\sigma_{c2}\sigma_{b2})\ldots 
i(1-2\sigma_{c(n-1)}\sigma_{b(n-1)})\sigma_{bn}\ ,\nonumber\\
&& 
a_{cn}=(-i)(1-2\sigma_{c1}\sigma_{b1})(-i)(1-2\sigma_{c2}\sigma_{b2})\ldots 
(-i)(1-2\sigma_{c(n-1)}\sigma_{b(n-1)})\sigma_{cn}\ ,\ \ \ \ldots
\eea
The factor $(1-2\sigma_{ci}\sigma_{bi})$ is $-1$ or $+1$ depending on 
whether the $i$-th location is occupied 
($\ua$) or not ($\da$): this means\ $(1-2\sigma_{ci}\sigma_{bi})^2=+1$ 
as can be checked explicitly as\ $(1-4\sigma_{ci}\sigma_{bi}+ 
4\sigma_{ci}\sigma_{bi}\sigma_{ci}\sigma_{bi})=1$.
Furthermore we see that the term 
\be
\big[\pm i(1-2\sigma_{ci}\sigma_{bi})\big]^\dag\ =\ 
\pm i(1-2\sigma_{ci}\sigma_{bi})\ ,
\ee
is hermitian as the $\pm i$ factors ensure, thereby ensuring that the 
$a_{bn}, a_{cn}$ operators are also hermitian: for instance
\be
a_{c2}^\dag = \sigma_{c2} [i(1-2\sigma_{b1}\sigma_{c1})] 
= -i(1-2\sigma_{c1}\sigma_{b1})\sigma_{c2} = a_{c2}\ .
\ee

Now it can be seen that the $a_b, a_c$-variables are anticommuting: \eg\
\bea
\{ a_{b2},\ a_{cn}\} &=& 
i(-i)^n\Big((1-2\sigma_{c1}\sigma_{b1})\sigma_{b2} (1-2\sigma_{c1}\sigma_{b1})
(1-2\sigma_{c2}\sigma_{b2})\ldots (1-2\sigma_{c(n-1)}\sigma_{b(n-1)})\sigma_{cn}
\nonumber\\ 
&& +\ (1-2\sigma_{c1}\sigma_{b1})(1-2\sigma_{c2}\sigma_{b2})\ldots 
(1-2\sigma_{c(n-1)}\sigma_{b(n-1)})\sigma_{cn} (1-2\sigma_{c1}\sigma_{b1})
\sigma_{b2} \Big)  \nonumber\\ 
&=& i(-i)^n \Big(
(1-2\sigma_{c1}\sigma_{b1})^2 \sigma_{b2} (1-2\sigma_{c2}\sigma_{b2})
\sigma_{cn} \ldots (1-2\sigma_{c(n-1)}\sigma_{b(n-1)}) \nonumber\\
&& +\ (1-2\sigma_{c1}\sigma_{b1})^2 (1-2\sigma_{c2}\sigma_{b2}) \sigma_{b2} 
\sigma_{cn} \ldots (1-2\sigma_{c(n-1)}\sigma_{b(n-1)}) \Big) \nonumber\\
&=& i(-i)^n \left( -\sigma_{b2} \sigma_{cn} + \sigma_{b2} \sigma_{cn} \right)
\ldots (1-2\sigma_{c(n-1)}\sigma_{b(n-1)})\ =\ 0
\eea
since the $\sigma$s at distinct locations are commuting. Similarly 
other anticommutation relations can be checked. The fact that the $a_{bn}$ 
contains a factor $i$ whereas $a_{cn}$ contains $-i$ ensures that the 
anticommutator works out correctly: \eg\ 
\bea
\{ a_{bn},\ a_{cn}\} &=& 
\{ \prod_{k=1}^{n-1} i(1-2\sigma_{ck}\sigma_{bk}) \sigma_{bn},\
\prod_{k=1}^{n-1} (-i)(1-2\sigma_{ck}\sigma_{bk}) \sigma_{cn} \} \nonumber\\
&=& i^{n-1} (-i)^{n-1} \{ \sigma_{bn},\ \sigma_{cn} \} = 1\ ,
\eea
where we have used the fact that each $(1-2\sigma_{ck}\sigma_{bk})$ factor 
commutes through the $\sigma_{bn}$ and $\sigma_{cn}$.\ 
Now note that 
\bea\label{JWbncn+1}
&& -J\sum_n \sigma_{b(n)} \sigma_{c(n+1)} %\nonumber\\
= -J\sum_n \left[ (1-2a_{c1}a_{b1})(1-2a_{c2}a_{b2})\ldots 
(1-2a_{c(n-1)}a_{b(n-1)}) a_{bn} \right]\times\quad \nonumber\\ 
&& \qquad\qquad\qquad\quad
\left[
(1-2a_{c1}a_{b1})(1-2a_{c2}a_{b2})\ldots (1-2a_{c(n-1)}a_{b(n-1)})
(1-2a_{cn}a_{bn}) a_{c(n+1)} \right] \nonumber\\
&& \qquad\qquad =
-J\sum_n \left[\prod (1-2a_{ci}a_{bi})^2\right]\ a_{bn} 
(1-2a_{cn}a_{bn}) a_{c(n+1)}\nonumber\\
&& \qquad\qquad = +J\sum_n a_{bn} a_{c(n+1)}\ .
\eea
We have used the fact that $a_{bn}$ commutes through each $(1-2a_{ci}a_{bi})$ 
factor, leaving a nontrivial action with $(1-2a_{cn}a_{bn})$.
It is now important to note that in the above calculation, we have 
assumed that the ghost-spin chain is infinite thereby allowing us to 
restrict to ``bulk'' terms: if the chain is finite, then there would 
be a boundary term of the form\ $\sigma_{bN}\sigma_{c1}$ which would 
not simplify to the above form (with the exception of 2 ghost-spins). 
For instance, for a finite chain of 3 ghost-spins, this boundary term 
gives\
$\sigma_{b3}\sigma_{c1} = (1-2a_{c1}a_{b1}) (1-2a_{c2}a_{b2}) a_{b3} a_{c1}$\
which simplifies to give a term of the form\ 
$-2a_{c2}a_{b2} a_{b3} a_{c1}$ which does not cancel with any other, and
is not of the above quadratic form.

\subsection{Ghost-spin chain for the $bc$-ghost CFT}

Consider a 1-dimensional ghost-spin chain with a nearest neighbour 
interaction Hamiltonian 
\be\label{Hgssigmabc}
H = J \sum_n \left( \sigma_{b(n)} \sigma_{c(n+1)} 
+ \sigma_{b(n)}\sigma_{c(n-1)} \right)\ ,
\ee
repeating (\ref{Hgssigmabc0}), 
where $n,\ n+1, n-1$ label nearest neighbour lattice sites in the chain, 
which comprises 2-state spin variables at each site. This is not quite 
Ising-like: in fact it describes a ``hopping'' type Hamiltonian, which 
kills an $\ua$-spin at site $n$ and creates it at site $n+1$, so that 
$\ua_n$ hops to $\ua_{n+1}$. It is useful to note that this Hamiltonian 
can also be written as  %(\ref{Hgssigmabc})
\be
H = J \sum_n \left( \sigma_{b(n)} \sigma_{c(n+1)} 
+ \sigma_{b(n+1)}\sigma_{cn} \right)\ ,
\ee
and so on a nearest neighbour pair $(n,n+1)$ the action of $H$ is 
seen quite generally to be
\bea
&& H |\ldots \ua_n\da_{n+1}\ldots\ran\ \ \xrightarrow{\ b_nc_{n+1}\ }\ \
|\ldots \da_n\ua_{n+1}\ldots\ran\ ,\nonumber\\
&& H |\ldots \da_n\ua_{n+1}\ldots\ran\ \ \xrightarrow{\ b_{n+1}c_n\ }\ \
|\ldots \ua_n\da_{n+1}\ldots\ran\ .
\eea

While (\ref{Hgssigmabc}) represents an 
infinite ghost-spin chain, it is worth illustrating its action by 
considering finite chains: so consider a system of two ghost-spin 
lattice sites, with
\be
H = J (\sigma_{b1} \sigma_{c2} + \sigma_{b2} \sigma_{c1})
\ee
where we have imposed periodic boundary conditions (which thus gives 
the second term). We then see that $H$ acts on the 4 states (in the 
commuting spin basis) as
\bea
H|\da\da\ran = 0 , && H|\ua\ua\ran = 0\ ,\nonumber\\
H|\da\ua\ran = J\sigma_{b2}\sigma_{c1}|\da\ua\ran = J|\ua\da\ran ,\
&& \  H|\ua\da\ran = J\sigma_{b1}\sigma_{c2}|\ua\da\ran = J|\da\ua\ran ,
\eea
since \eg\ $\sigma_{b1}\sigma_{c2}$ kills $|\da\ua\ran$ and so on.\ 
The energy expectation values (for states with nonzero norm 
$\lan\psi|\psi\ran$) are
\be
\lan E\ran = {\lan\psi|H|\psi\ran\over \lan\psi|\psi\ran}\ ;\qquad\quad
e^1_\pm: \ \lan E\ran = 0\ \ [{\rm ground\ states}];\qquad 
e^2_\pm: \ \lan E\ran = J\ \ [{\rm excited\ states}]\ ,
\ee
using the basis in (\ref{2gsbasis:e12+-}), \ie\ 
$e^1_\pm={1\over\sqrt{2}}(|\da\da\ran\pm|\ua\ua\ran)$ etc.
This gives the partition function\
$Z = \sum_{s_i} e^{-\beta E[s_i]}\ =\ 2 \left( 1 + e^{-\beta J} \right)$\
which is identical to that for 2 ordinary spins.\ 
Consider now 3 lattice sites (again with periodic boundary conditions): 
the Hamiltonian is
\be
H = J (\sigma_{b1}\sigma_{c2} + \sigma_{b2}\sigma_{c1} 
+ \sigma_{b2}\sigma_{c3} + \sigma_{b3}\sigma_{c2} 
+ \sigma_{b3}\sigma_{c1} + \sigma_{b1}\sigma_{c3})\ .
\ee
The action of $H$ on the 8 states is
\bea
H|\da\da\da\ran = 0\ , && H|\ua\ua\ua\ran = 0\ ,\nonumber\\
|\da\ua\da\ran \xrightarrow{\ b_2c_1 + b_2c_3\ }\ \ |\ua\da\da\ran 
+ |\da\da\ua\ran\ ; && 
|\da\da\ua\ran \xrightarrow{\ b_3c_1 + b_3c_2\ }\ \ |\ua\da\da\ran 
+ |\da\ua\da\ran\ ; \nonumber\\
|\ua\ua\da\ran \xrightarrow{\ b_1c_3 + b_2c_3\ }\ \ |\da\ua\ua\ran 
+ |\ua\da\ua\ran\ ; &&
|\ua\da\ua\ran \xrightarrow{\ b_1c_2 + b_3c_2\ }\ \ |\da\ua\ua\ran 
+ |\ua\ua\da\ran\ ; \nonumber\\
|\ua\da\da\ran \xrightarrow{\ b_1c_2 + b_1c_3\ }\ \ |\da\ua\da\ran 
+ |\da\da\ua\ran\ ; &&
|\da\ua\ua\ran \xrightarrow{\ b_2c_1 + b_3c_1\ }\ \ |\ua\da\ua\ran 
+ |\ua\da\da\ran\ .
\eea
Thus some $H$ eigenstates with norms $\pm 1$ are
\be
|\da\da\da\ran\pm|\ua\ua\ua\ran\ ;\qquad 
(|\da\ua\da\ran + |\da\da\ua\ran + |\ua\da\da\ran) \pm 
(|\ua\da\ua\ran + |\ua\ua\da\ran + |\da\ua\ua\ran)\ .
\ee
Norms for some generic states then are
\be
\al_1 (|\da\ua\da\ran + |\da\da\ua\ran + |\ua\da\da\ran) \pm \al_2
(|\ua\da\ua\ran + |\ua\ua\da\ran + |\da\ua\ua\ran)\ \ \longrightarrow\ \
3 (\al_1^*\al_2 + \al_2^*\al_1)
\ee
\be
\al_1 |\da\da\da\ran + \al_2 |\ua\ua\ua\ran\ \ \longrightarrow\ \ 
2 (\al_1^*\al_2 + \al_2^*\al_1)
\ee
while their energy eigenvalues are $0$ and $2\al_1\pm 2\al_2$ respectively.

\vspace{3mm}

\noindent {\bf \emph{gs $\ra$ bc:}}\ \ Starting with the ghost-spin chain 
Hamiltonian in the commuting spin variables
\be\label{Hsigmabncn}
H = +J \sum_n \left( \sigma_{b(n)} \sigma_{c(n+1)} 
+ \sigma_{b(n)}\sigma_{c(n-1)} \right)\ ,
\ee
we see using the Jordan-Wigner transformation (\ref{JWsigmabcnabcn}), 
(\ref{JWabcnsigmabcn}), and the simplification (\ref{JWbncn+1}), that the 
Hamiltonian simplifies as
\bea
H &=& J\sum_n\ \Big( i^{n-1} [1][2]\ldots[n-1] a_{bn} 
(-i)^{n} [1][2]\ldots[n] a_{c(n+1)}\ \nonumber\\
&&\qquad\qquad
+\ i^{n-1} [1][2]\ldots[n-1] a_{bn} (-i)^{n-2} [1][2]\ldots[n-2] a_{c(n-1)} 
\Big)\ ,
\eea
where\ $[k]\equiv (1-2a_{ck}a_{bk})$.\ Commuting the various $[k]$ 
factors gives
\bea\label{Habncn}
H &=& J \sum_n \Big( (-i) a_{bn} (1-2a_{cn}a_{bn}) a_{c(n+1)}\ +\ 
i (1-2a_{c(n-1)}a_{b(n-1)}) a_{bn} a_{c(n-1)} \Big) \nonumber\\
&=& i J a_{bn} \left( a_{c(n+1)} - a_{c(n-1)} \right)\ .
\eea
In what follows, we will take a continuum limit of this system where 
$J$ is scaled as $J\sim {1\over 2a}$: then we see that the difference 
becomes the derivative, \ie\ $iJa_{bn}(a_{c(n+1)}-a_{c(n-1)})\ra -b\del c$, 
in the continuum limit.
Note that the Hamiltonian (\ref{Habncn}) is hermitian: after 
anticommuting the $a_{bn}$ through, we have\ 
$H^\dag = (-i)(a_{c(n+1)}-a_{c(n-1)})a_{bn}=H$.\ The ghost-spin Hamiltonian 
(\ref{Hgssigmabc}) that we began with was also hermitian of course.

\noindent {\bf \emph{Momentum space variables:}}\ \ 
So far we have been working with the lattice variables, which are real
space representation of the spin degrees of freedom.  To give the
momentum space description of these operators let us consider
the Fourier transform of the real space operators
\be
a_{bn} = {1\over\sqrt{N}} \sum_k e^{-ikn} b_k\ ,\qquad 
a_{cn} = {1\over\sqrt{N}} \sum_k e^{-ikn} c_k\ .
\ee
The hermiticity of $a_{bn}, a_{cn}$ imposes a relation between
negative Fourier modes and hermitian conjugate operators,
\be
b_{-k}=b_k^\dag\ ,\qquad c_{-k}=c_k^\dag\ .
\ee
The inverse transforms are
\be
b_k = {1\over\sqrt{N}} \sum_n e^{ikn} a_{bn}\ ,\quad\ \
c_k = {1\over\sqrt{N}} \sum_n e^{ikn} a_{cn}\ ,\qquad {\rm with}\quad
{1\over N} \sum_n e^{in(k+k')} = \delta_{k+k',0}\ .
\ee
The operators $a_{bn}, a_{cn}$ are fermionic and satisfy
anticommutation relations.  The anticommutation relation between then
translates into the following anticommutation relations between $b$ and
$c$ Fourier modes,
\be
\{ b_k,\ c_{k'} \} = {1\over N} \sum_{n,m} e^{ikn+ik'm} \{ a_{bn},\ a_{cm} \}
= \delta_{k+k',0}\ ,\qquad \{ b_k,\ b_{k'} \} = 0\ ,\quad 
\{ c_k,\ c_{k'} \} = 0\ .
\ee
In addition to these modes we see that there also exist ``zero mode''
operators, which are momentum space analogs of the centre of mass
modes, 
\be\label{b0c0}
k=0:\qquad b_0 = {1\over\sqrt{N}} \sum_n a_{bn}\ ,\quad
c_0 = {1\over\sqrt{N}} \sum_n a_{cn}\qquad\Rightarrow\qquad
\{ b_0,\ c_0 \} = 1\ .
\ee
Note that we are considering a chain of $N$ fermionic ghost-spins with 
$N$ odd and the momentum moding
\be
k = \pm m {2\pi\over N} \equiv k_{m\pm}\ , \qquad m=1,2,\ldots, {N-1\over 2}
\qquad [N\ odd],
\ee
which in the large $N$ continuum limit becomes $k\ra [-\pi,\pi]$.
To illustrate this, consider 3 ghost-spins: the lattice sites are 
labelled by $n=0,1,2$, and $k=-{2\pi\over 3}, 0, {2\pi\over 3}$,\ giving
\bea
&& b_{\pm 2\pi/3} = {1\over\sqrt{3}} \left( e^{\pm i(2\pi/3)(0)} a_{b0} + 
e^{\pm i(2\pi/3)(1)} a_{b1} + e^{\pm i(2\pi/3)(2)} a_{b2} \right)\ ,\nonumber\\
&& b_0={1\over \sqrt{3}} (a_{b0} + a_{b1} + a_{b2})\ ,
\eea
and likewise for the $c_k$ operators. These Fourier modes allow a 
faithful mapping of the various spin states in terms of the momentum 
basis. The anticommutation relations are
\be
\{b_{\pm 2\pi/3}, c_0\} = {1\over 3} (\{a_{b0},a_{c0}\} + e^{\pm i2\pi/3} 
\{a_{b1},a_{c1}\} + e^{\pm i4\pi/3} \{a_{b2},a_{c2}\} ) = 
{1\over 3} (1+\omega_3 + \omega_3^2) = 0\ ,
\ee
with $\omega_3=e^{\pm i2\pi/3}$ a 3rd root of unity.
Likewise for general odd $N$, the anticommutation relation vanishes as\
$\{b_{k}, c_0\} = {1\over N} \sum_{n=0}^{N-1} e^{i(2\pi m/N) n} = 
{1\over N} (1 + \omega_N + \ldots + \omega_N^{N-1})=0$\ 
with $\omega_N=e^{i(2\pi m/N)},\ m=\pm 1,2,\ldots,{N-1\over 2}$, a general 
$N$-th root of unity.   It is worth pointing out at this stage that
for $N$ even, it turns out that the zero mode  
operators, if they exist, do not yield sensible anticommutation relations 
with the other modes. This is perhaps due to implicit anti-periodic 
boundary conditions. Our description of these momentum modes here with 
$N$ odd is similar to the discussion in \eg\ \cite{Kogut:1979wt}.

There is a pair of ground states for these momentum basis modes defined 
by the zero modes $b_0, c_0$, in (\ref{b0c0}),
\be
b_0|\da_{bc}\ran=0\ ,\quad b_0|\ua_{bc}\ran=|\da_{bc}\ran\ ,\quad
c_0|\da_{bc}\ran=|\ua_{bc}\ran\ ,\quad c_0|\ua_{bc}\ran=0\ ,
\ee
and all higher modes $b_k, c_k$, with $k>0$ annihilate $|\da_{bc}\ran, 
|\ua_{bc}\ran$.\ Note that these are distinct from the position basis 
states described earlier. Then states such as $|\da_{bc}\ran-|\ua_{bc}\ran$
clearly have negative norm. Excited states such as 
$(b_{-k}-c_{-k})(|\da_{bc}\ran-|\ua_{bc}\ran)$\ also have negative norm\ 
$(\lan\da_{bc}-|\ua_{bc}\ran)|(b_k-c_k)
(b_{-k}-c_{-k})(|\da_{bc}\ran-|\ua_{bc}\ran)=-2$ using the $b_k,c_k$ 
oscillator algebra.

In terms of the momentum basis modes, the ghost-spin chain Hamiltonian 
becomes
\bea
H &=& J {i\over N} \sum_n\sum_{k,k'} e^{-ikn} b_k \big( e^{-ik'(n+1)} c_{k'} 
- e^{-ik'(n-1)} c_{k'} \big)\nonumber\\
&=& 
iJ \sum_{k,k'} b_k c_{k'}\ \delta_{k+k',0}  \big( e^{-ik'} - e^{+ik'} \big)
= 2J \sum_k \sin k'\ b_k c_{k'}\ \delta_{k+k',0}\ .
\eea
Reinstating the lattice spacing $a$ by replacing $k$ by $ka$ in the
sine function and then taking the continuum limit $a\ra 0$ gives
\be
H = 2J \sum_k \sin (ka) b_{-k} c_k\quad\longrightarrow\
2Ja \sum_k k b_{-k}c_k\ .
\ee
In order to obtain the critical theory we need to scale the coupling $J$ 
as $J\sim {1\over 2a}$ while taking the continuum limit to obtain a 
nonzero finite expression as $a\ra 0$: this is simply a way of ensuring 
that the nearest neighbour lattice interaction leads to nontrivial 
continuum interactions as the lattice spacing goes to zero. This then 
gives
\be\label{Hgsbc}
H = \sum_{k>0} k \left( b_{-k}c_k + c_{-k}b_k \right) + \zeta\ .
\ee
The constant $\zeta$ is a normal ordering constant that arises as usual 
after rewriting creation operators to the left of annhilation operators.
%where the second term has been obtained from 
%$\sum_{k>0} (-k) b_k c_{-k}$ after using $b_kc_{-k}+c_{-k}b_k=1$.

The $H$ above is of the same form as $L_0$ of the $bc$-ghost CFT with 
$c=-2$.  We can construct other Virasoro generators by picking up
appropriate Fourier modes of the ghost-spin chain Hamiltonian density 
$ia_{bn} (a_{c(n+1)}-a_{c(n-1)})$.  For example
\begin{equation}
  \label{eq:1}
  L_n = \sum_k (n-k)  b_{k}c_{n-k}\ .
\end{equation}
%  L_n = \sum_{k>0} (k+n) \left( b_{-k}c_{k+n} + c_{-k}b_{k+n} \right)\ .
Thus in the continuum limit we recover conformal invariance and we can
express the Virasoro generators in terms of modes of $b$ and $c$ ghosts.
In addition to the Virasoro symmetry we also have the ghost current
symmetry $ J_g(z) =\ :\!cb\!:\!(z)$ .

It is worth asking what the symmetries of the original ghost-spin 
chain Hamiltonian (\ref{Hgssigmabc}) were. In this regard we note that 
$H$ term-by-term respects a phase rotation symmetry
\be
\sigma_{b(n)}\ra e^{i\al} \sigma_{b(n)}\ ,\qquad
\sigma_{c(n+1)} \ra e^{-i\al} \sigma_{c(n+1)}\ .
\ee
This is a microscopic reflection of the $U(1)$ symmetry in the continuum 
$bc$-CFT. In addition, note that there is a global scaling symmetry 
\be\label{gschainScalingsymm}
a\ra \xi^{-1} a\ ,\quad H\ra \xi H\ ,\qquad 
\sigma_{b(n)}\ra \xi^\lambda \sigma_{b(n)}\ ,\qquad
\sigma_{c(n+1)} \ra \xi^{1-\lambda} \sigma_{c(n+1)}\ .
\ee
We see that the ghost-spin variables $(\sigma_b,\sigma_c)$ exhibit 
this symmetry for any constant $\lambda$\ (although $\lambda=1$ was 
implicit in most of our discussion above): this is the reflection of 
the fact that the $bc$-CFT is a conformal theory for any conformal 
weights $(h_b, h_c) = (\lambda,1-\lambda)$. This arises from the fact 
that each term in $H$ involves two separate variables allowing a 
partial ``cancellation'' of the scaling factor $\xi$. This would not 
be possible for an Ising-like Hamiltonian, \eg\ of the form
(\ref{2gsIsingHint}), (\ref{NgsIsingHint}).

Further let us recall that for a general $bc$-CFT with weights 
$(h_b, h_c) = (\lambda,1-\lambda)$, the energy-momentum tensor is\ 
$T(z)=:(\del b)c: - \lambda \del (:bc:) 
= -:b\del c: + (1-\lambda) \del (:bc:)$. This can be rewritten as\
$T(z)_\lambda = - \lambda :b\del c: + (1-\lambda) :(\del b)c:$~.\ It 
is then useful to note that the lattice discretization of the last 
expression is
\bea
&& \lambda J \sum_n i a_{bn} \left( a_{c(n+1)} - a_{c(n-1)} \right)
- (1-\lambda) J \sum_n i \left( a_{b(n+1)} - a_{b(n-1)} \right) a_{cn}
\nonumber\\
&& \qquad\qquad 
\longrightarrow\ \ \sum_n iJ a_{bn} \left( a_{c(n+1)} - a_{c(n-1)} \right) ,
\eea
where we have taken\ 
$iJa_{bn}(a_{c(n+1)}-a_{c(n-1)})\ra -b\del c$ from (\ref{Habncn}), and 
the last simplification can be seen by appropriately recasting 
the $\sum_n$ in the second infinite lattice sum in the first line. In 
other words, the local expression $iJa_{bn} (a_{c(n+1)}-a_{c(n-1)})$
can be split into the two terms in $T(z)_\lambda$ for any $\lambda$. 
This is consistent with the fact that $T(z)_\lambda$ is $-b\del c$
apart from a total derivative. Thus the lattice Hamiltonian
(\ref{Habncn}) obtained from (\ref{Hgssigmabc}) captures the general
$bc$-CFT$_\lambda$ equally well in the continuum limit $a\ra 0$ with
$J\sim {1\over 2a}$ , along with the scaling (\ref{gschainScalingsymm}).

We have thus argued that the ghost-spin chain with Hamiltonian 
(\ref{Hgssigmabc}) with weights $(\lambda,1-\lambda)$ for the 
ghost-spin variables $\sigma_{bn}, \sigma_{cn}$, under the scaling 
symmetry (\ref{gschainScalingsymm}) maps to the $bc$-ghost CFT with 
conformal weights $(h_b, h_c) = (\lambda,1-\lambda)$ in the continuum 
limit.  Note that while the scaling symmetry can be demonstrated in both
the ghost-spin variables as well as the fermionic ghost-spin variables 
$a_{bn}, a_{cn}$ representation, the Jordan-Wigner transformation which 
is a non-local relation between these two representations does not
have this scaling symmetry. For ghost fields with conformal weights
$(h_b, h_c) = (\lambda,1-\lambda)$, the Virasoro generators are given
by\ $ L_n = \sum_k (n\lambda -k) b_{k}c_{n-k}$, and the normal ordering 
constant $\zeta$ in (\ref{Hgsbc}) above is fixed by the Virasoro 
algebra of the $L_n$s of the $bc$-CFT as usual.

%\vspace{10mm}

\section{Discussion}

We have studied 1-dimensional chains of ghost-spins with nearest
neighbour interactions amongst them, developing the description of
ghost-spins in \cite{Narayan:2016xwq,Jatkar:2016lzq}. Ghost-spins,
2-state spin variables with indefinite norm, serve as simple quantum
mechanical toy models for theories with negative norm states. In the
finite ghost-spin chains, we have described how the Ising-like nearest
neighbour interaction helps organize and clarify the study of
entanglement earlier and we have further developed the properties of
the reduced density matrix and its entanglement entropy. We have then
studied a family of infinite ghost-spin chains with hopping type
Hamiltonian, where defining fermionic ghost-spin variables through a
Jordan-Wigner transformation maps these ghost-spin chains in the
continuum limit to the $bc$-ghost CFTs. It may be interesting to
explore other ghost-like field theories in this light and more
generally the space of non-unitary CFTs that ghost-spin ensembles
provide microscopic realizations for.

Along the lines of the Ising-like ghost-spin chains, 
a simple generalization of an infinite ghost-spin chain is the 
transverse Ising model for a ghost-spin chain
with Hamiltonian \eg\ $H = -J\sum_i (s^z_is^z_{i+1} + g\sum_i
s^x_i)$,\ where $s^z$ are the ghost-spin variables $s$ we have been
describing so far, and $s^x$ are complementary variables (not
commuting with $s^z$).  For $g\sim 0$, the ground states are $s^z$
eigenstates $|+^N\rangle,\ |-^N\rangle$, as discussed previously,
while for $g\gg 1$, the ground state is the $\sigma^x$ eigenstate
$|\da\ran$: this is very similar to the ordinary transverse
Ising spin chain, except that the variables here represent ghost-spins
with indefinite norms and thus encode negative norm states. It
would seem that $g=1$ is a critical point where some scale invariant
theory emerges. In light of our discussion here on the $bc$-ghost CFT
which arises from a very different ghost-spin chain, it is unclear
what this critical theory might be.\\
Another interesting system is a ``ghost-spin glass'', with a
Hamiltonian of the Ising spin glass form but with $N$ ghost-spins\ $H
= -\sum J_{ij} s_is_j$\ with $i,j=1,\ldots,N$. The couplings $J_{ij}$
are not restricted to nearest neighbour and so represent random
nonlocal interaction couplings.  In the $\{\pm\}$-basis, it would
appear based on the discussions in sec.~3 that this system would have
parallels with ordinary spin-glasses (see \eg\ \cite{Denef:2011ee} 
for a relatively recent review), exhibiting many nearly degenerate 
ground states, but also containing negative norm states. 
It would be interesting to explore these.

The appearance of the $bc$-ghost system in the continuum limit of the
infinite ghost-spin chain points towards a gauge symmetry which has
been fixed using the Faddeev-Popov method.  Such a symmetry would
become manifest if this ghost system is coupled to ordinary matter.
In familiar theories with gauge symmetry, the negative norm sector
decouples from any physical process, a truncation which is technically
implemented by the familiar BRST procedure.  In the present case also,
we expect that an appropriate BRST symmetry will enable a truncation
of the full indefinite norm Hilbert space to the physical Hilbert space
which comprises positive norm states alone, thereby leading in 
principle to positive entanglement entropy. We hope to report on 
this in the future.

The original motivation for constructing ``ghost-spins'' in
\cite{Narayan:2016xwq} was to explore solvable toy models for
ghost-CFTs and study their entanglement properties: this builds on
earlier studies \cite{Narayan:2015vda, Narayan:2015oka} of 
generalizations of the Ryu-Takayanagi formulation \cite{Ryu:2006bv,HRT}
to gauge/gravity duality for de Sitter space or $dS/CFT$
\cite{Strominger:2001pn,Witten:2001kn,Maldacena:2002vr}.  In
\cite{Narayan:2015vda,Narayan:2015oka}, the areas of certain complex
codim-2 extremal surfaces (involving an imaginary bulk time
parametrization) were found to have structural resemblance with
entanglement entropy of dual Euclidean CFTs, effectively equivalent to
analytic continuation from the Ryu-Takayanagi expressions in
$AdS/CFT$. In $dS_4$ the areas are real and negative. Certain attempts
were made in \cite{Narayan:2016xwq} towards gaining some insight on
this in CFT and quantum mechanical toy models: certain 2-dim
ghost-CFTs under certain conditions were found to yield negative
entanglement entropy using the replica formulation
\cite{Calabrese:2004eu}. Likewise a toy model of two ghost-spins was
found to yield the reduced density matrix and associated entanglement
properties reviewed earlier in sec.~\ref{sec:revsGs}.
In the context of $dS/CFT$
\cite{Strominger:2001pn,Witten:2001kn,Maldacena:2002vr}, de Sitter
space is conjectured to be dual to a hypothetical Euclidean
non-unitary CFT that lives on the future boundary ${\cal I}^+$, with
the dictionary $\Psi_{dS}=Z_{CFT}$\ \cite{Maldacena:2002vr}, where
$\Psi_{dS}$ is the late-time wavefunction of the universe with
appropriate boundary conditions and $Z_{CFT}$ the dual CFT partition
function.  This usefully organizes de Sitter perturbations,
independent of the actual existence of the CFT. The dual CFT$_d$
energy-momentum tensor correlators reveal central charge coefficients
${\cal C}_d\sim i^{1-d}{R_{dS}^{d-1}\over G_{d+1}}$ in $dS_{d+1}$
(effectively analytic continuations from $AdS/CFT$). This is real and
negative in $dS_4$ so that $dS_4/CFT_3$ is reminiscent of ghost-like
non-unitary theories. In \cite{Anninos:2011ui}, a higher spin $dS_4$
duality was conjectured involving a 3-dim CFT of anti-commuting
$Sp(N)$ (ghost) scalars (studied previously in
\cite{LeClair:2006kb,LeClair:2007iy}). In this light, we are thinking
of ensembles of ghost-spins as toy models for the latter $Sp(N)$
theories and thereby $dS_4$ possibly. In general such an ensemble of a
large number $N$ of ghost-spins is non-unitary, containing large
families of negative norm states. However as we have seen, there are
subsectors of positive norm states as well, which in fact appear
perfectly well-defined and sensible. It is interesting to speculate on
possible parallels in the context of a possible dual cosmology.

%$PT$-symmetric formulation may have interesting implications for the
%dS/CFT correspondence but we will that line here.

\vspace{10mm}  %\newpage
{\footnotesize \noindent {\bf Acknowledgements:}\ \ It is a pleasure to 
thank Dionysios Anninos and Ashoke Sen for some early discussions.
We thank the Organizers of the ISM 2016 Indian Strings Meeting, IISER Pune, 
for hospitality while this work was in progress.
KN thanks the String Theory Groups at HRI, Allahabad, and TIFR Mumbai 
as well as the organizers of ``String Theory: Past and Present'' 
Discussion Meeting (SpentaFest), ICTS Bangalore, for hospitality 
while this work was in progress. The work of KN is partially supported by 
a grant to CMI from the Infosys Foundation and of DPJ by the DAE project 
12-R\&D-HRI-5.02-0303.}

\vspace{3mm}

%\appendix

\end{document}